\documentclass[aps,prd,superscriptaddress,nofootinbib,tighten,preprint]{revtex4}
\usepackage[utf8]{inputenc}
\usepackage[english]{babel}
\usepackage{amsmath}
\usepackage{subfigure}
\usepackage{cancel}
\usepackage{amsfonts}
\usepackage{amssymb}
\usepackage{graphicx}
\usepackage{xcolor}
\newcommand{\be}{\begin{equation}}
\newcommand{\ee}{\end{equation}}
\newcommand{\bea}{\begin{eqnarray}}
\newcommand{\eea}{\end{eqnarray}}

\newcommand{\zmt}{Z_{BL}}

\newcommand{\nn}{\nonumber}

\catcode`@=12

\begin{document}

\title{WIMP-FIMP option and neutrino masses \vspace{0.05cm}\\ via a novel anomaly-free $B-L$ symmetry}

\author{Sarif Khan}
\email{sarifkhan@cau.ac.kr}
\affiliation{Department of Physics, Chung-Ang University, Seoul 06974, Korea.}
\author{Hyun Min Lee}
\email{hminlee@cau.ac.kr}
\affiliation{Department of Physics, Chung-Ang University, Seoul 06974, Korea.}

\section*{}

\vspace{1cm}

\begin{abstract} 
We propose a novel $U(1)_{B-L}$  model with singlet dark matter fermions composed of WIMP and FIMP, which is anomaly-free without a need for introducing right-handed neutrinos. Fermion dark matter masses are generated after the $U(1)_{B-L}$ is broken spontaneously, so the Yukawa couplings for WIMP and FIMP components can be distinguished by the hierarchical values of the vacuum expectation values of the single scalar fields.  Moreover, the $U(1)_{B-L}$ gauge boson receives a TeV-scale mass for a tiny extra gauge coupling, so it goes out of equilibrium from the rest of the model content in the early Universe. 
Both the  $U(1)_{B-L}$ gauge boson and FIMP component are produced from
the decays of the bath particles, and the former can decay into FIMP DM and/or WIMP DM before BBN. The WIMP component can reside in the resonance region of the Higgs bosons or dominantly annihilate into a pair of singlet-like scalars. Thus, there is a flexibility to choose a small mixing between the visible and dark sectors, thereby evading all the current direct and indirect detection bounds.
Furthermore, we show that WIMP and FIMP components can coexist in suitable fractions, depending on the choice of model parameters, allowing for additional protection for WIMP DM against various experimental bounds. Finally, we identify the dimension-6 and dimension-7 operators for Majorana neutrino masses in our model, being consistent with the $U(1)_{B-L}$ gauge symmetry, and provide a possibility of extending the model with additional singlet fermions for neutrino masses.

\end{abstract}
\maketitle

\section{Introduction}
\label{Intro}
When dark matter (DM) mostly self-annihilates into the Standard Model (SM) sector, most of the parameter space for the so-called Weakly Interacting Massive Particle (WIMP) has been explored with the increased detector volume and state-of-the-art DM detection techniques.  
However, there is no evidence for the DM interactions with the SM apart from gravitational interaction. So far, the experiments have looked for DM with a weak coupling to the SM, but with null results. For instance, the recent data from LUX-ZEPLIN 
\cite{LZCollaboration:2024lux} have ruled out the spin-independent direct detection (SIDD) cross-section up to $\sigma_{SI} \sim 10^{-48}$ ${\rm cm}^{2}$ for a DM mass of about 50 GeV. With such a severe bound, Higgs portal DM can only survive near the resonance region. Therefore, we need to consider different kinds of DM candidates with weaker couplings to the SM or go beyond the standard scenarios such as WIMP DM annihilating into an additional dark sector. 

There are still many possibilities to overcome the SIDD bounds and open up larger parameter spaces that have not been explored yet. One popular alternative to evade these bounds is the freeze-in DM 
scenario \cite{McDonald:2001vt, Hall:2009bx}, referred to in the literature as Feebly Interacting Massive Particle (FIMP), where the coupling strength is so tiny that freeze-in DM is very difficult to probe in ongoing experiments. Another possibility is multi-component DM 
\cite{Abdallah:2019svm, Costa:2022oaa, Covi:2022hqb, Belanger:2022gqc, Costa:2022lpy, Khan:2024biq}, 
which is very well motivated because our visible world consists of many particles, from quarks/leptons to gauge bosons as force carriers.
In the case of multi-component DM, we can consider scenarios with 
both WIMP-type DM, both FIMP-type DM or a combination of WIMP and FIMP DM. For multicomponent WIMP DM, one WIMP DM candidate could annihilate into another WIMP DM candidate, and the second WIMP DM candidate could annihilate into the visible sector. By suitably choosing the fraction of the second WIMP DM candidate, we can always evade the existing bounds, and the freeze-out mechanism will practically remain effective for a longer time, even if the DD experiments reach the 
neutrino floor \cite{OHare:2021utq}.
As discussed, it is challenging to detect FIMP DM, but a detectable FIMP DM can be achieved, with an appropriate adjustment of the reheating temperature \cite{Cosme:2023xpa, Silva-Malpartida:2023yks, Arcadi:2024wwg, Lee:2024wes}. 

In this article, we explore the possibility of WIMP and FIMP DM and search for the parameter space that has not been explored by any experiments. To this purpose, we consider a well-known $U(1)_{B-L}$ gauge extension of the SM, but without three right-handed neutrinos which used to be required to cancel the gauge anomalies \cite{Biswas:2016ewm}. The main disadvantage of $U(1)_{B-L}$ gauge symmetry with right-handed neutrinos 
\cite{Mohapatra:1980qe, Georgi:1981pg, Wetterich:1981bx, Lindner:2011it, Okada:2010wd} 
is that there is no suitable candidate for DM unless one of the right-handed neutrinos is detached from the visible sector through the symmetry or it is assumed to have tiny interactions with the visible sector \cite{Abdallah:2019svm}.

In our work, we introduce four additional chiral fermions
\cite{Patra:2016ofq, Biswas:2018yus} 
with suitable charges to cancel the gauge anomalies arising from $[SM]^2 
\times U(1)_{B-L}$ and $[U(1)_{B-L}]^3$. 
In Refs. \cite{Patra:2016ofq, Biswas:2018yus}, the authors studied single-component WIMP DM\footnote{In Ref. \cite{Patra:2016ofq},
one of the extra Dirac fermions and beyond SM scalars were considered heavy and did not contribute to the DM phenomenology. In contrast, our study considers the full particle setup in the DM phenomenology.} near the resonance region. In the present work, we explore a multi-component DM consisting of WIMP and FIMP combinations.

To cancel the gauge anomalies introduced by $U(1)_{B-L}$, the SM is extended with four chiral fermions, comprising two left-handed and two right-handed fermions with fractional charges, along with two singlet scalars. One CP-odd scalar {\it d.o.f.} among the two becomes the longitudinal component of the additional $U(1)_{B-L}$ gauge boson, rendering it massive, while the other becomes a physical CP-odd particle that actively participates in the DM phenomenology \cite{Kim:2024cwp}.

In the present model, there are three CP-even scalars: two arising from the singlet scalars and one from the SM Higgs doublet. All three scalars contribute to the DM phenomenology. The $U(1)_{B-L}$ charge assignments of the singlet scalars are made such that two Dirac fermions can be realized from the four chiral fermions. Both Dirac fermions act as DM candidates, with one being of the WIMP type ($\psi_1$, as defined in Section \ref{model}) and the other of the FIMP type ($\psi_2$, as defined in Section \ref{model}). Moreover,
we take the mixing among two Dirac fermions to zero by making the associated
Yukawa couplings to zero or introducing two independent $Z_2$ symmetries. 
The distinction between the couplings for WIMP and FIMP DM  arises from the hierarchy between the VEVs of the singlet scalars, $v_1$ and $v_2$, namely,  $v_1 \gg v_2$. However, in the scenarios where $v_1 \sim v_2$ and $v_{1,2} \sim \mathcal{O}(\text{TeV})$, both DM candidates are of the WIMP type, whereas for $v_{1,2} \sim \mathcal{O}(\text{PeV})$, both are of the FIMP type. We consider a GeV-scale mass for the $Z_{BL}$ gauge boson, which places the extra gauge coupling $g_{BL}$ in the feeble regime for a PeV-scale VEV of $\phi_1$. In the presence of such a feeble coupling rendering $Z_{BL}$ non-thermal, we determine the non-thermal distribution for $Z_{BL}$ from the first principles and compare it with the results that would have been obtained with a thermal distribution.

In the case of WIMP-type DM, aside from the Higgs resonance region, there is a large parameter space where DM predominantly annihilates into BSM scalars. This provides the freedom to arbitrarily choose the mixing angle between the visible and dark sectors, effectively evading the existing bounds for WIMP DM. On the other hand, for FIMP-type DM candidates, we primarily consider the decay modes for DM production, which dominate over the annihilation contributions.

In our study, we include all the relevant constraints on the model, such as perturbativity bounds, vacuum stability bounds, Higgs measurements, direct detection, indirect detection, oblique parameters, and Big Bang nucleosynthesis (BBN). Moreover, we also consider higher dimensional operators for Majorana neutrino masses being consistent with the $U(1)_{B-L}$ symmetry. We consider the possibility of realizing such higher dimensional operators by introducing three right-handed fermions with suitable charges while keeping the gauge anomalies cancelled and letting them interact with lepton doublets.

The rest of the paper is organized as follows. In Section \ref{model}, we describe our model in detail. Section \ref{constraints} discusses all the constraints associated with our study. In Section \ref{result}, we present our findings on DM. Finally, in Section \ref{conclusion}, we conclude our work. There is one appendix containing the analytical formulas for decay widths and collision terms, used in our study.

\section{The model}
\label{model}

We present the model setup in the extension of the SM with the $U(1)_{B-L}$ gauge symmetry. 
In this section, we discuss the conditions for gauge anomalies in our model and the mass spectrum and the mixing angles in the scalar sector after the $U(1)_{B-L}$ symmetry is broken. In the basis for physical fermions and scalars, we obtain the mass spectrum for singlet fermions and their interaction terms. We also show how neutrino masses are obtainable in the effective theory with $U(1)_{B-L}$ and the extension with three right-handed neutrinos.

\begin{center}
\begin{table}[h!]
\begin{tabular}{||c|c|c|c||}
\hline
\hline
\begin{tabular}{c}
    Gauge\\
    Group\\ 
    \hline
    
    ${\rm SU(2)}_{\rm L}$\\  
    \hline
    ${\rm U(1)}_{\rm Y}$\\ 
    \hline
    $U(1)_{B-L}$\\ 
\end{tabular}
&

\begin{tabular}{c|c|c}
    \multicolumn{3}{c}{Quarks}\\ 
    \hline
    $Q_{L}^{i}=(u_{L}^{i},d_{L}^{i})^{T}$&$u_{R}^{i}$&$d_{R}^{i}$\\ 
    \hline
    $2$&$1$&$1$\\ 
    \hline
    $1/6$&$2/3$&$-1/3$\\ 
    \hline
    $1/3$&$1/3$&$1/3$\\ 
\end{tabular}
&
\begin{tabular}{c|c}
    \multicolumn{2}{c}{Leptons}\\
    \hline
    $L_{L}^{i}=(\nu_{L}^{i},e_{L}^{i})^{T}$ & $e_{R}^{i}$\\
    \hline
    $2$&$1$\\
    \hline
    $-1/2$&$-1$\\
    \hline
    $-1$&$-1$\\
\end{tabular}
&
\begin{tabular}{c}
    \multicolumn{1}{c}{Higgs doublet}\\
    \hline
    $\phi_{h}$\\
    \hline
    $2$\\
    \hline
    $1/2$\\
    \hline
    $0$\\
\end{tabular}\\
\hline
\hline
\end{tabular}
\caption{Charge assignments for the SM matter content under the SM and $U(1)_{B-L}$ gauge groups. }
\label{tab1}
\end{table}
\end{center}

The complete Lagrangian takes the following form,
\begin{eqnarray}
\mathcal{L}& = &\mathcal{L}_{SM} + 
\sum_{i =1, 2} \left( D_{\mu} \phi_{i} \right)^{\dagger} \left( D_{\mu} \phi_{i} \right) 
+ \mathcal{L}^{Kin}_{BL}
 - \mathcal{V}\left(\phi_{h},\phi_{1}, \phi_{2} \right)\,.   
\end{eqnarray}
On the other hand, the kinetic terms and the Yukawa terms for the additional fermions can be written as
follows,
\begin{eqnarray}
\mathcal{L}^{Kin}_{BL} &=& \sum_{X = \xi_{1L}, \xi_{2L}, \xi_{1R}, \chi_{2R}} \bar{X} i \cancel{D} X + \alpha_{1} \bar \xi_{1L} \chi_{1R} \phi_{2} 
+ \alpha_{2} \bar \xi_{2L} \chi_{2R} \phi_{1} 
+ \beta_{1} \bar \xi_{2L} \chi_{1R} \phi_{1} \nonumber \\ && 
+ \beta_{2} \bar \xi_{1L} \chi_{2R} \phi_{2} + {\it h.c.}
\label{Yukawa-exotic-fermion}
\end{eqnarray}
where $\cancel{D} X\equiv \gamma^{\mu} D_{\mu} X$. The covariant derivatives are $D_{\mu} X = \partial_{\mu} X - i g_{BL} n^{X}_{BL} Z_{BL} X$ where $g_{BL}$ is the gauge coupling, $n^{X}_{BL}$ is the $U(1)_{B-L}$ 
charge of the field $X$, as shown in Tables \ref{tab1} and \ref{tab2}, and $Z_{BL}$ is the gauge field of the $U(1)_{B-L}$ symmetry.

 \begin{center}
\begin{table}[h!]
\begin{tabular}{||c|c|c||}
\hline
\hline
\begin{tabular}{c}
    Gauge\\
    Group\\ 
    \hline
    
    ${\rm SU(2)}_{\rm L}$\\  
    \hline
    ${\rm U(1)}_{\rm Y}$\\ 
    \hline
    $U(1)_{B-L}$\\ 
\end{tabular}
&

\begin{tabular}{c|c|c|c}
    \multicolumn{4}{c}{Extra fermions}\\ 
    \hline
    $\xi_{1L}$&$\xi_{2L}$&$\chi_{1L}$ &$\chi_{2L}$ \\ 
    \hline
    $1$&$1$&$1$&$1$\\ 
    \hline
    $0$&$0$&$0$&$0$\\ 
    \hline
    $a$&$b$&$c$ &$c$\\ 
\end{tabular}
&
\begin{tabular}{c|c}
    \multicolumn{2}{c}{Extra scalars}\\
    \hline
    $\phi_{1}$ & $\phi_{2}$\\
    \hline
    $1$ & $1$ \\
    \hline
    $0$ & $0$\\
    \hline
    $n$ & $2 n$\\
\end{tabular}\\
\hline
\hline
\end{tabular}
\caption{Charge assignments for extra fermions and scalar fields under the SM and $U(1)_{B-L}$ gauge groups.}
\label{tab2}
\end{table}
\end{center}

From the Lagrangian in Eq. (\ref{Yukawa-exotic-fermion}), we have the possibility to explore many types of DM scenarios by choosing different values for $\alpha_{1,2}$ and $\beta_{1,2}$.
First, if we have $\alpha_{1,2} \sim \beta_{1,2} \sim \mathcal{O}(0.1)$, we can realize one or two-component WIMP-type DM, depending on whether the decay of one to another is open. The recent direct detection bound from LUX-ZEPLIN \cite{LZCollaboration:2024lux} 
has tightly constrained the parameter space for 
WIMP-type DM candidates if they dominantly annihilate into the SM sector.
On the other hand, if we consider $\alpha_{1,2} \sim \beta_{1,2} \sim \mathcal{O}(10^{-10})$, then we can have FIMP-type DM candidates, which will be safe from all kinds of bounds but would be hardly probed in the near future with the ongoing experiments. The aforementioned Yukawa couplings for WIMP and FIMP type DM can be taken for $v_{1,2}$ of TeV and PeV scales,  respectively, for GeV scale DM masses.

In the following discussion, we focus on a combination of WIMP and FIMP-type DM candidates, which remain safe from all the existing bounds while being detectable in the ongoing experiments. Additionally, due to the rich particle content, we can have a regime where WIMP-type DM mainly annihilates into the dark sector, i.e., singlet-like scalars. This gives us the freedom to choose the mixing angle between the dark sector and the visible sector freely without affecting the DM production, allowing us to evade all the terrestrial bounds.
Therefore, in the present work, we consider $\alpha_{1} \sim \mathcal{O}(0.1)$ and $\alpha_{2} \sim \mathcal{O}(10^{-10})$, which can be easily achieved by choosing hierarchical\footnote{In the literature, hierarchical VEVs between scalar multiplets under the same gauge group have been considered for obtaining 
suitable phenomenology, 
{\it e.g.}, in the Type II seesaw mechanism \cite{Konetschny:1977bn,
Schechter:1980gr}, where one can take the 
triplet scalar VEV to the eV scale, leading to the doubly 
charged Higgs predominantly decaying into 
two leptons \cite{Chun:2019hce, Dev:2019hev}.}
VEVs, $v_{1} \gg v_{2}$. 
We can ensure the stability for the two-component DM scenario either by introducing two independent $Z_2\times Z'_2$ symmetries for WIMP and FIMP  or by taking $\beta_{1,2}=0$  in Eq. (\ref{Yukawa-exotic-fermion}) (or zero mixing angles between the singlet fermions, i.e. $\theta_{L,R} = 0$, defined in Eq. (\ref{WIMP-FIMP-mixing})).

The scalar potential for the SM Higgs doublet and two singlet scalars obeying the complete gauge symmetry takes the following,
\begin{eqnarray}
\mathcal{V}(\phi_{h},\phi_{1},\phi_{2} ) & = &
- \mu^{2}_{h} \left( \phi^{\dagger}_{h} \phi_{h} \right) +
\lambda_{h} \left( \phi^{\dagger}_{h} \phi_{h} \right)^{2}
- \mu^{2}_{1} \left( \phi^{\dagger}_{1} \phi_{1} \right) +
\lambda_{1} \left( \phi^{\dagger}_{1} \phi_{1} \right)^{2}
- \mu^{2}_{2} \left( \phi^{\dagger}_{2} \phi_{2} \right) 
\nonumber \\
&+&
\lambda_{2} \left( \phi^{\dagger}_{2} \phi_{2} \right)^{2} + \lambda_{h1} \left( \phi^{\dagger}_{h} \phi_{h} \right)
\left( \phi^{\dagger}_{1} \phi_{1} \right)
+ \lambda_{h2} \left( \phi^{\dagger}_{h} \phi_{h} \right)
\left( \phi^{\dagger}_{2} \phi_{2} \right) \nonumber \\
&+& \lambda_{12} \left( \phi^{\dagger}_{1} \phi_{1} \right)
\left( \phi^{\dagger}_{2} \phi_{2} \right)
+ \mu \left( \phi_{2} \phi^{\dagger\,2}_{1} 
+ \phi^{\dagger}_{2} \phi^{2}_{1} \right)
\end{eqnarray}

\subsection{Gauge anomalies}

In the present work, we have introduced singlet fermions, which are charged under the $U(1)_{B-L}$ gauge symmetry. 
We have chosen the charges of such singlet fermions in such a way that they make non-trivial contributions to
$\left[ U(1)_{B-L} \right]^{3}$ and 
$\left[{\rm Gravity}\right]^{2} \times U(1)_{B-L}$. Moreover, we also have
constraints on the charges of the particles involved in the
Yukawa terms. Therefore, the constraints can be summarised in the following
manner,
\begin{eqnarray}
 \left[ U(1)_{B-L} \right]^{3} &\rightarrow& a^{3} + b^{3} - 2 c^{3} = 3\,,
 \\ \nonumber 
\left[{\rm Gravity} \right]^{2} \times U(1)_{B-L} &\rightarrow&
a + b - 2c = 3\,, \\ \nonumber
{\rm Yukawa\,\,terms} &\rightarrow& a-c = 2 n \,\,\,{\rm and}\,\,\,
b-c = n\,.  
\end{eqnarray}     
Once we solve the above set of equations together, we have only two
different choices for the $B-L$ charges,
\begin{eqnarray}
\left(a,b,c,n\right) = \left(1,0,-1,1 \right) \,\,{\rm and}\,\, 
\left(\frac{4}{3}, \frac{1}{3}, -\frac{2}{3},1 \right)\,. 
\end{eqnarray}
In the following, we choose the option for the fractional charges 
$\left(\frac{4}{3}, \frac{1}{3}, -\frac{2}{3},1 \right)$
to continue our phenomenology. 
{If we consider the other choices with integer charges, we open up the Majorana mass terms for them and the Yukawa interaction terms with the lepton doublet, which will open up decay modes, thereby eliminating the DM candidate. This scenario is similar to the case for the $U(1)_{B-L}$ gauge symmetry with three right-handed neutrinos cancelling the gauge anomalies, which can decay into the SM sector very 
easily \cite{Biswas:2016ewm}.}

\subsection{Scalar sector}
In the present work, we have one scalar doublet and two singlet scalars, so we expand them, after the symmetry breaking, as follows,
\begin{eqnarray}
\phi_{h} =
\begin{pmatrix}
G^{+} \\
\frac{v+h + i G^{0}}{\sqrt{2}} 
\end{pmatrix},\quad
\phi_{1} = \frac{v_{1} + H_{1} + i A_{1}  }{\sqrt{2}},\quad
\phi_{2} = \frac{v_{2} + H_{2} + i A_{2}  }{\sqrt{2}}.
\end{eqnarray}
After the symmetry breaking, we consider the tadpole conditions with respect to $h, H_1, H_2$, and take the values of other fields equal to zero, to get
\begin{eqnarray}
\mu^2_{h} &=& \frac{1}{2} \left[ 2 \lambda_{h} v^2_{h} + \lambda_{h1} v^2_{1} + \lambda_{h2} v^2_{2} \right], \nonumber \\
\mu^2_{1} &=& \frac{1}{2} \left[ 2 \lambda_{1} v^2_{1} + \lambda_{h1} v^2_{h} + v_{2} \left( 2 \sqrt{2} \mu + \lambda_{12} v_{2} \right) \right],
\nonumber \\
\mu^2_2 &=& \frac{1}{2} \left[ 2 \lambda_{2} v^2_{2} 
+ \lambda_{h2} v^2_{h} + \lambda_{12} v^2_{1} + \frac{\sqrt{2} \mu v^2_{1}}{v_{2}}  \right]\,.
\end{eqnarray}  

From the second derivative of the scalar potential with respect to the fields and the tadpole conditions, we can have the neutral Higgs mass matrix in the basis $(h\,\,H_{1}\,\,H_{2})$,
\begin{eqnarray}
M^2_{scalar} = 
\begin{pmatrix}
2 \lambda_{h} v^2_{h} & \lambda_{h1} v_{h} v_{1} & \lambda_{h2} v_{h} v_{2} \\
\lambda_{h1} v_{h} v_{1} & 2 \lambda_{1} v^2_{1} 
& v_{1} \left( \sqrt{2} \mu + \lambda_{12} v_{2} \right) \\
\lambda_{h2} v_{h} v_{2} & v_{1} \left( \sqrt{2} \mu + \lambda_{12} v_{2} \right) & \left( -\frac{\mu v^2_{1}}{\sqrt{2} v_{2}} + 2 \lambda_{2} v^2_{2} \right)
\end{pmatrix}.
\end{eqnarray}
We can diagonalise the above neutral Higgs mass matrix by the unitary 
matrix and the mass eigenstates, $(h_{1}\,\,h_{2}\,\,h_{3})$, can be introduced 
in the following way,
\begin{eqnarray}
\begin{pmatrix}
h\\
H_{1}\\
H_{2}
\end{pmatrix}
= U_{ij} 
\begin{pmatrix}
h_{1}\\
h_{2}\\
h_{3}
\end{pmatrix}
\end{eqnarray} 
where the unitary matrix $U_{ij}\,(i,j=1,2,3)$ can be expressed as
\begin{eqnarray}
U_{ij} =
\begin{pmatrix}
c_{12} c_{13} & s_{12} c_{13} & s_{13} \\
-s_{12} c_{23} - c_{12} s_{23} s_{13} & c_{13} c_{23} - s_{12} s_{23} s_{13}
& s_{23} c_{13} \\
s_{12} s_{23} - c_{12} c_{23} s_{13} & -c_{12} s_{23} - s_{12} c_{23} s_{13}
& c_{23} c_{13}
\end{pmatrix}\,.
\label{U-PMNS}
\end{eqnarray}

In the same way, we can write down the mass matrix for the CP odd eigenstates in the basis $(A_{1}\,\,A_{2})$, as follows,
\begin{eqnarray}
M^2_{CP-odd} =
\begin{pmatrix}
 - 2 \sqrt{2} \mu v_{2} & \sqrt{2} \mu v_{1} \\
 \sqrt{2} \mu v_{1} & - \frac{\mu v_{1}}{\sqrt{2} v_{2}}
\end{pmatrix}.
\end{eqnarray}
There is also a Goldstone boson associated with the SM $Z$ boson is independent of the other CP odd states.
We can diagonalise the CP-odd mass matrix and get the eigenvalues as
\begin{eqnarray}
M^2_{A} = -2 \sqrt{2} \mu v_{2} \left( 1 + \frac{v^2_{1}}{4 v^2_{2}} \right)\,,\quad M_{G_{BL}} = 0\,.
\end{eqnarray}
Here, the mass eigenstates for the  CP odd eigenstates are obtained in the following way,
\begin{eqnarray}
\begin{pmatrix}
G_{BL} \\
A
\end{pmatrix}
=
\begin{pmatrix}
 \cos\beta & -\sin\beta \\
 \sin\beta & \cos\beta
\end{pmatrix}
\begin{pmatrix}
A_{1} \\
A_{2}
\end{pmatrix}
\end{eqnarray}
where $\sin\beta = \frac{1}{\sqrt{1 + \frac{v^2_{1}}{4 v^2_{2}}}}$,
$\cos\beta = \frac{v_{1}}{2 v_{2}\sqrt{  1 + \frac{v^2_{1}}{4 v^2_{2}}  }}$ and $\tan\beta = \frac{2 v_{2}}{v_{1}}$\,.
In the above notation, $G_{BL}$ is associated with the Goldstone boson
for the $U(1)_{B-L}$ gauge boson whose mass is given by
\begin{eqnarray}
M^2_{Z_{BL}} &=& g^2_{BL} v^2_{1} + 4 g^2_{BL} v^2_{2}\,\nonumber \\
 &=& g^2_{BL} v^2_{1} \left(1 + \tan^{2}\beta \right)
\end{eqnarray}

\subsection{Singlet fermion masses and interactions}

Once the singlet scalars take the VEVs, we can write down the mass term 
for the BSM fermions as follows,
\begin{eqnarray}
\mathcal{L}_{\xi \chi} = \begin{pmatrix}
\bar \xi_{1\,L} & \bar \xi_{2\,L}
\end{pmatrix}
\begin{pmatrix}
\frac{\alpha_{1} v_{2}}{\sqrt{2}} & \frac{\beta_{2} v_{2}}{\sqrt{2}} \\
\frac{\beta_{1} v_{1}}{\sqrt{2}} & \frac{\alpha_{2} v_{1}}{\sqrt{2}} 
\end{pmatrix}
\begin{pmatrix}
\chi_{1 R} \\
\chi_{2 R}
\end{pmatrix}
+ {\it h.c.}
\end{eqnarray} 
We can relate the singlet fermions to the left-handed and right-handed fermions of  two Dirac mass eigenstates, $\psi_{1,2} = \psi_{1,2\,L} \oplus \psi_{1,2\,R}$, by
\begin{eqnarray}
\begin{pmatrix}
\xi_{1L}\\
\xi_{2L}
\end{pmatrix}
= U_{L} 
\begin{pmatrix}
\psi_{1L}\\
\psi_{2L}
\end{pmatrix}\,,
\,\,\,
\begin{pmatrix}
\chi_{1R}\\
\chi_{2R}
\end{pmatrix}
= U_{R} 
\begin{pmatrix}
\psi_{1R}\\
\psi_{2R}
\end{pmatrix}\,,\,\,{\rm with}\,\,
U_{L,R} = \begin{pmatrix}
\cos\theta_{L,R} & \sin\theta_{L,R}\\
-\sin\theta_{L,R} & \cos\theta_{L,R}
\end{pmatrix}\,.
\label{WIMP-FIMP-mixing}
\end{eqnarray} 
Therefore, using the above field redefinitions, we can write down 
\begin{eqnarray}
\begin{pmatrix}
M_{1} & 0 \\
0 & M_{2}
\end{pmatrix}
= U^{T}_{L} \begin{pmatrix}
\frac{\alpha_{1} v_{2}}{\sqrt{2}} & \frac{\beta_{2} v_{2}}{\sqrt{2}} \\
\frac{\beta_{1} v_{1}}{\sqrt{2}} & \frac{\alpha_{2} v_{1}}{\sqrt{2}} 
\end{pmatrix} U_{R}
\end{eqnarray}
where $M_{1,2}$ are the masses of the Dirac fermions.\,. 
The fermionic mixing angles $\theta_{L,R}$
can be represented such that $\theta_{L,R} \rightarrow 0$ as $\beta_{1,2} \rightarrow 0$, and
the limits are still true for $\alpha_{2} \ll 1$. 
The fermionic mixing angles, $\theta_{L,R}$, can be
written in terms of the Yukawa couplings, $\alpha_{1,2}$ and $\beta_{1,2}$,
and physical fermion masses, $M_{1,2}$, in the following way,
\begin{eqnarray}
\tan\theta_{R} &=& \frac{M_{1} v_{2} \beta_{2} + M_{2} v_{1} \beta_{1} }
{M_{2} v_{1} \alpha_{2} - M_{1} v_{2} \alpha_{1}}\,, \nonumber \\
\tan\theta_{L} &=& \frac{M_{1}}{M_{2}} 
\frac{\alpha_{1} \tan\theta_{R} + \beta_{1}}{\alpha_{1} - \beta_{2} \tan\theta_{R}} 
\,.
\end{eqnarray}
We see from the above relation that if we fix the $\theta_{R}$ then $\theta_L$
is automatically fixed with different values.
Finally, the Yukawa terms associated with the BSM fermions $\psi_{1,2}$ 
and the Higgs bosons, $h_{1,2,3}$, take the following form,
\begin{eqnarray}
\mathcal{L}^{Yuk}_{\psi} &=&
\sum_{i=1,2,3} \alpha_{11i} \bar \psi_{1L} \psi_{1R} h_{i} +
\sum_{i=1,2,3} \alpha_{12i} \bar \psi_{1L} \psi_{2R} h_{i} +  
\sum_{i=1,2,3} \alpha_{21i} \bar \psi_{2L} \psi_{1R} h_{i} 
\nonumber \\ &+&
\sum_{i=1,2,3} \alpha_{22i} \bar \psi_{2L} \psi_{2R} h_{i} 
+ \mathrm{\it i}\, \alpha_{11A}\, \bar \psi_{1L} \psi_{1R} A
+ \mathrm{\it i}\, \alpha_{12A}\, \bar \psi_{1L} \psi_{2R} A 
+ \mathrm{\it i} \, \alpha_{21A}\, \bar \psi_{2L} \psi_{1R} A
\nonumber \\
&+& \mathrm{\it i}\, \alpha_{22A}\, \bar \psi_{2L} \psi_{2R} A + {\it h.c.}\,.
\end{eqnarray}
where the Yukawa interactions for physical states take the following form,
\begin{eqnarray}
\alpha_{11i} &=& \frac{M_{1}}{\sqrt{2} v_{1} v_{2}} 
\left[ U_{3i} v_{1} + U_{2i} v_{2} + \left( U_{3i} v_{1} - U_{2i} v_{2} \right) \cos 2 \theta_{L} \right]\,, \nonumber \\
\alpha_{12i} &=& \frac{\sqrt{2} M_{2}}{v_{1} v_{2}} 
\left[  \left( U_{3i} v_{1} - U_{2i} v_{2} \right) \cos \theta_{L}
\sin\theta_{L} \right]\,, \nonumber \\
\alpha_{21i} &=& \frac{\sqrt{2} M_{1} }{ v_{1} v_{2} } 
\left[  \left( U_{3i} v_{1} - U_{2i} v_{2} \right) \cos\theta_{L}
\sin\theta_{L} \right]\,, \nonumber \\
\alpha_{22i} &=& \frac{M_{2}}{\sqrt{2} v_{1} v_{2}} 
\left[ U_{3i} v_{1} + U_{2i} v_{2} + \left( -U_{3i} v_{1} + U_{2i} v_{2} \right) \cos 2 \theta_{L} \right].\,
\label{psipsihi}
\end{eqnarray}
Here, the matrix $U$ is defined in Eq. (\ref{U-PMNS}), and  $U_{3A} = \cos\beta$ and $U_{2A} = \sin\beta$ for $i=A$\,.

In the basis of the mass eigenstates $\psi_{1,2}$, we can have the $U(1)_{B-L}$ gauge interaction terms with the  gauge boson $Z_{BL}$, as follows,
\begin{eqnarray}
\mathcal{L}_{\psi Z_{BL}} &=& - \frac{g_{BL}}{3} 
\biggl[ \bar \psi_{1} \gamma^{\mu} \left( (3 \cos^{2}\theta_{L} +1)P_{L}
- 2 P_{R} \right) \psi_{1}
+ \bar \psi_{2} \gamma^{\mu} \left( (3 \sin^{2}\theta_{L} +1)P_{L}
- 2 P_{R} \right) \psi_{2} \nonumber \\
&+& \bar \psi_{1} \gamma^{\mu} (2 \sin^{2}\theta_{L}) P_{L} \psi_{2}
+ \bar \psi_{2} \gamma^{\mu} (2 \sin^{2}\theta_{L}) P_{L} \psi_{1}
 \biggr] Z_{BL\,\mu}\,.
\end{eqnarray}

For the later discussion, we consider $\psi_1$ as WIMP-type DM and $\psi_2$ as FIMP DM, with the associated couplings in the freeze-out and freeze-in regimes, respectively. This can be achieved easily when $v_{1} \gg v_{2}$ and all the BSM particle masses are in the sub-TeV range. Then, we can see that the extra gauge coupling $g_{BL}$ and the Yukawa couplings $\alpha_{22i}$ fall in the feeble regime whereas the Yukawa couplings $\alpha_{11i}$ are in the freeze-out regime.
If $\theta_L=0$, a stable two-component dark matter composed of WIMP and FIMP is possible. Otherwise, the heavier one will decay into the lighter one, leaving only one component DM.

\subsection{Neutrino masses}

For the particle content with the $U(1)_{B-L}$ gauge symmetry, the neutrino masses can be generated by introducing an additional triplet scalar, as has been 
studied in Ref. \cite{Patra:2016ofq} or by introducing the right-handed neutrinos 
through inverse seesaw mechanism Ref. \cite{Biswas:2018yus}.
 In our model, we anticipate higher-dimensional operators for Majorana neutrino masses that are $U(1)_{B-L}$ gauge invariant,
\begin{eqnarray}
\mathcal{L}_{Neutrino} = \kappa_{ij}\frac{ \left( L_{i} \phi_{h}\right)
\left( L_{j} \phi_{h}\right)}{\Lambda} \frac{\phi^2_{1}}{\Lambda^{2}} 
+ \kappa^{\prime}_{ij} \frac{ \left( L_{i} \phi_{h}\right) \left( L_{j} \phi_{h}\right)}{\Lambda} \frac{\phi_{2}}{\Lambda} + {\it h.c.}\,. 
\end{eqnarray}  
where $\kappa_{ij}, \kappa^{\prime}_{ij}$ ($i,j = 1,2,3$) are 
the coefficients and $\Lambda$ is the cut-off scale.
The higher-dimensional terms can be originated from the extension of the current set-up with three additional right-handed fermions that are singlets under the SM gauge group but charged under the $U(1)_{B-L}$ in such a way that the gauge anomalies remain cancelled. 
With these additional fermions, we can realize the Type-I seesaw mechanism 
for the generation of light active neutrino masses. 
In Table \ref{tab3}, we show the $U(1)_{B-L}$ charge assignment for the additional right-handed singlets which does not affect the gauge anomaly condition.   

 \begin{center}
\begin{table}[h!]
\begin{tabular}{||c|c||}
\hline
\hline
\begin{tabular}{c}
    Gauge\\
    Group\\ 
    \hline
       $SU(2)_{L}$\\   
    \hline 
    $U(1)_{B-L}$\\  
    \hline 
\end{tabular}
&
\begin{tabular}{c|c|c}
    \multicolumn{3}{c}{Fermionic Fields}\\ 
    \hline
    $N_{1}$&$N_{2}$&$N_{3}$  \\ 
\hline
    $1$&$1$&$1$\\    
    \hline
    $1$&$-1$&$0$\\ 
    \hline
\end{tabular}\\
\hline
\hline
\end{tabular}
\caption{Right-handed singlet fermions and their corresponding
charges under $SU(2)_{L}$ and $U(1)_{B-L}$ gauge groups. All the 
additional fermions have zero hypercharges.}
\label{tab3}
\end{table}
\end{center}

The Lagrangian for the additional right-handed fermions up to dimension-5 operators, 
which are allowed under SM and $U(1)_{B-L}$ gauge groups,
can be written as
\begin{eqnarray}
\mathcal{L}_{N} &=& y_{e1} \bar{L}_{e} \tilde{\phi}_{h} N_{1} \frac{\phi_{2}}{\Lambda}
+ y_{e2} \bar{L}_{e} \tilde{\phi}_{h} N_{2} + y_{e3} \bar{L}_{e} 
\tilde{\phi}_{h} N_{3} \frac{\phi_{1}}{\Lambda} 
+ y_{\mu 1} \bar{L}_{\mu} \tilde{\phi}_{h} N_{1} \frac{\phi_{2}}{\Lambda}
+ y_{\mu 2} \bar{L}_{\mu} \tilde{\phi}_{h} N_{2} \nonumber \\
& +& y_{\mu 3} \bar{L}_{\mu} \tilde{\phi}_{h} N_{3} \frac{\phi_{1}}{\Lambda}
+ y_{\tau 1} \bar{L}_{\tau} \tilde{\phi}_{h} N_{1} \frac{\phi_{2}}{\Lambda}
+ y_{\tau 2} \bar{L}_{\tau} \tilde{\phi}_{h} N_{2} 
+ y_{\tau 3} \bar{L}_{\tau} \tilde{\phi}_{h} N_{3} \frac{\phi_{1}}{\Lambda}
+ Y_{11} N_{1} N_{1} \phi_{2} \nonumber \\
& +& Y_{12} N_{1} N_{2} \phi_{2} + Y_{13} N_{1} N_{3} \phi_{2}
+ Y_{22} N_{2} N_{2} \phi_{2} + Y_{23} N_{2} N_{3} \phi_{1} + M_{33} N_{3} N_{3}
+ {\it h.c.}\,.
\end{eqnarray} 
Once the $U(1)_{B-L}$ symmetry is spontaneously broken, we can write the neutrino 
mass matrix in the $(\nu^{c}_{L,i}\,\,\,N_{i})^{T}$ basis ,as follows,
 \begin{eqnarray}
\mathcal{L}_{N-mass} = 
\begin{pmatrix}
\bar \nu^{c}_{L\,i} & \bar N_{i}
\end{pmatrix}
 \begin{pmatrix}
 0 & m_{D} \\
 m^{T}_{D} & M_{R}
 \end{pmatrix}
 \begin{pmatrix}
 \nu_{L\,i}\\
 N^c_{i}
 \end{pmatrix}
 + {\it h.c.}
 \end{eqnarray}
 where the Dirac ($m_{D}$) and Majorana ($M_{R}$) mass matrices are,
 \begin{eqnarray}
 m_{D} =
 \begin{pmatrix}
 \frac{y_{e1} v v_{2}}{2 \Lambda} & \frac{y_{e2} v}{\sqrt{2}} & 
 \frac{y_{e3} v v_{1}}{2 \Lambda} \\
  \frac{y_{\mu 1} v v_{2}}{2 \Lambda} & \frac{y_{\mu 2} v}{\sqrt{2}}
  & \frac{y_{\mu 3} v v_{1}}{2 \Lambda} \\
  \frac{y_{\tau 1} v v_{2}}{2 \Lambda} & \frac{y_{\tau 2} v }{\sqrt{2}}
  & \frac{y_{\tau 3} v v_{1}}{2 \Lambda}
 \end{pmatrix},\,\,\,
 M_{R} = \begin{pmatrix}
 \frac{Y_{11} v_{2}}{\sqrt{2}} & 
 \frac{Y_{12} v_{2}}{\sqrt{2}} & \frac{Y_{13} v_{1}}{\sqrt{2}} \\
  \frac{Y_{12} v_{2}}{\sqrt{2}} & 
 \frac{Y_{22} v_{2}}{\sqrt{2}} & \frac{Y_{23} v_{1}}{\sqrt{2}} \\
  \frac{Y_{13} v_{1}}{\sqrt{2}} & 
 \frac{Y_{23} v_{1}}{\sqrt{2}} & M_{33}
 \end{pmatrix}\,.
\end{eqnarray}   
Once we diagonalize the above neutrino mass matrix in the limit with $M_{R} \gg m_{D}$, we can write down the light neutrino mass matrix ($m_{\nu}$) and the heavy neutrino mass matrix ($M_{N}$) as,
\begin{eqnarray}
m_{\nu} \simeq - m^{T}_{D} M^{-1}_{R} m_{D}\,,\,\,\,M_{N} \simeq M_{R}
\end{eqnarray}
Then, we can use the above mass matrices to get the neutrino masses and the correct oscillation parameters easily \cite{Esteban:2024eli}. For example, if we consider the masses of the right-handed fermions to be $M_{N,ij} \sim \text{TeV}$, we only have to take the Dirac mass masses to be  $m_{D,ij} \sim 10^{-4}$ GeV to obtain the light neutrino masses of order $0.01$ eV.
There are free parameters in $M_{R}$ and $m_{D}$, $y_{fi}$ and $Y_{ij}$ ($f = e,\mu,\tau$ and $i,j = 1,2,3$) and $\Lambda$, which give us the freedom to make each component fall within the suitable range, even if $v_{1}$ and $v_{2}$ are hierarchical. Moreover, if we demand that the cut-off scale is higher than $v_{1,2}$ and $v_{1} \gg v_{2}$, then $m_{D,11}, m_{D,21}, m_{D,31}$ become subdominant, but we still have enough free parameters to account for the neutrino oscillation parameters. 

\section{Constraints}
\label{constraints}

In this section, we discuss various constraints on the model, such as perturbativity bounds, vacuum stability bounds for theoretical consistency, and experimental bounds such as  DM relic density, DM direct and indirect detection bounds, collider bounds (mainly from Higgs decays), bounds from BBN and oblique parameters.

\subsection{Perturbativity bounds}

The quartic couplings in the scalar potential can be written in terms of the masses and the mixing angles for the scalar fields, as follows,
\begin{eqnarray}
\lambda_{h} &=& \frac{U^2_{11} M^2_{h_{1}} + U^2_{12} M^2_{h_{2}} 
+ U^2_{13} M^2_{h_{3}} }{2 v^2_{h}}\,,\nonumber \\
\lambda_{1} &=& \frac{U^2_{21} M^2_{h_{1}} + U^2_{22} M^2_{h_{2}} 
+ U^2_{23} M^2_{h_{3}} }{2 v^2_{1}}\,, \nonumber \\
\lambda_{2} &=& \frac{U^2_{31} M^2_{h_{1}} + U^2_{32} M^2_{h_{2}} 
+ U^2_{33} M^2_{h_{3}} - M^2_{A} \cos^{2}\beta  }{2 v^2_{2}} \,,\nonumber\\
\lambda_{h1} &=& \frac{U_{11} U_{21} M^2_{h_{1}} 
+ U_{12} U_{22} M^2_{h_{2}} + U_{13} U_{23} M^2_{h_{3}} }{v_{h}v_{1} }
\,, \nonumber \\
\lambda_{h2} &=& \frac{U_{11} U_{31} M^2_{h_{1}} 
+ U_{12} U_{32} M^2_{h_{2}} + U_{13} U_{33} M^2_{h_{3}} }{v_{h}v_{2} }
\,, \nonumber \\
\lambda_{12} &=& \frac{U_{21} U_{31} M^2_{h_{1}} 
+ U_{22} U_{32} M^2_{h_{2}} + U_{23} U_{33} M^2_{h_{3}} + M^2_{A}
\sin\beta\, \cos\beta }{v_{1}v_{2} }
\,. \nonumber \\
\label{quartic-coupling-expression}
\end{eqnarray}
In the following analysis, we assume that all the quartic couplings are in the perturbative regime, namely, between $0$ and $4 \pi$.

\subsection{Vacuum stability bounds}

The bound from the below of the potential for the higher values of the fields can be
expressed in terms of the quartic couplings in the following way,
\begin{eqnarray}
&&\lambda_{h} > 0\,,\,\, \lambda_{1} > 0\,,\,\, \lambda_{2} > 0\,,\,\,
 \lambda_{12} + 2 \sqrt{\lambda_{1} \lambda_{2} } > 0\,,
 - 2 \sqrt{\lambda_{h} \lambda_{1}} \leq \lambda_{h1} \leq 2 \sqrt{\lambda_{h} \lambda_{1}}\,,\nonumber \\
&& - 2 \sqrt{\lambda_{h} \lambda_{2}} \leq \lambda_{h2} \leq 2 \sqrt{\lambda_{h} \lambda_{2}}\,,
2 \lambda_{h} \lambda_{12} - \lambda_{h1} \lambda_{h2} 
+ \sqrt{\left( 4 \lambda_{h} \lambda_{1} -\lambda^2_{h1}  \right)
\left( 4 \lambda_{h} \lambda_{2} -\lambda^2_{h2}  \right)} \geq 0\,.
\end{eqnarray}

\subsection{DM relic density}
In our model, we can account for the Planck constraints on the total DM relic density, incorporating the contributions from both the WIMP and FIMP DM components. Specifically, we consider a $7\sigma$ variation in the DM relic density obtained from Planck 2018 \cite{Planck:2018vyg}, as detailed below,
\begin{eqnarray}
0.1116 \leq \left( \Omega_{\psi_{1}} + \Omega_{\psi_{2}} \right) h^{2} \leq 0.1284\,.
\end{eqnarray}

\subsection{DM direct detection bounds}
\begin{figure}[h!]
\centering
\includegraphics[angle=0,height=4.5cm,width=14.5cm]{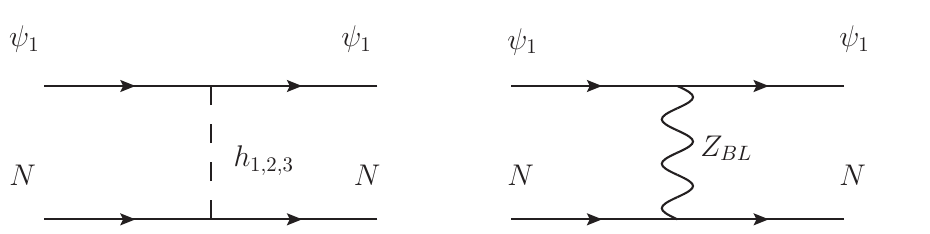}
\caption{Direct detection processes relevant in our study.} 
\label{DM-DD}
\end{figure}

The direct detection cross-section for the DM-nucleus scattering process,
$\psi_{1} N \rightarrow \psi_{1} N$ (as shown in Fig. \ref{DM-DD}) 
can be expressed in the non-relativistic limit, as follows,
\begin{eqnarray}
\sigma_{\psi_{1}} = \frac{\mu^{2}}{\pi} \left[ \frac{f_{N} M_{N}}{v} 
\sum_{i = 1,2,3} \frac{U_{1i} \alpha_{11i}}{M^2_{h_{i}}} 
+ \frac{f_{Z_{BL}} g^{2}_{BL} \left( 3 \cos^{2}\theta_{L} -1 \right) }{18 M^2_{Z_{BL}}} \right]^{2}
\label{DD-expression}
\end{eqnarray} 
where $\mu = \frac{M_{\psi_{1}} M_{N}}{M_{\psi_{1}} + M_{N}}$ and 
$f_{N} \sim 0.3$ \cite{Cline:2013gha}, $f_{\zmt} = 3$ \cite{Belanger:2008sj}. In our study, we consider the multi-component DM scenario, so we need to multiply by the fraction of $\psi_{1}$ DM component, $f_{\psi_{1}}$, for the DD direction bounds,
when compared with the recent LUX-ZEPLIN data \cite{LZCollaboration:2024lux}.

\subsection{DM indirect detection bounds}

\begin{figure}[h!]
\centering
\includegraphics[angle=0,height=4.5cm,width=14.5cm]{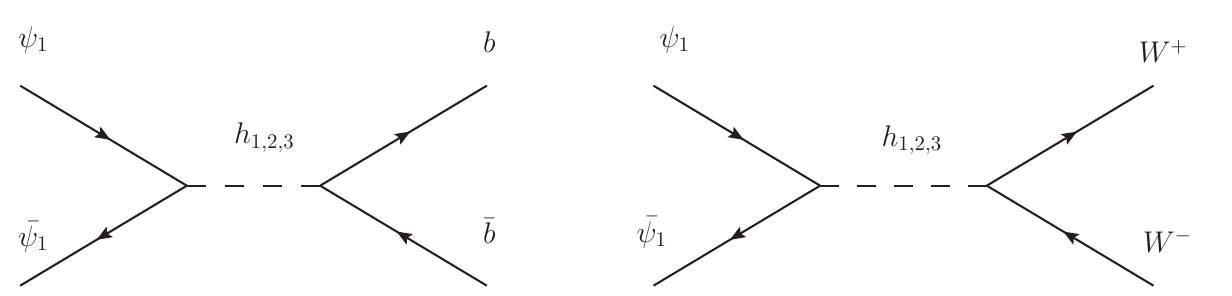}
\caption{Indirect detection processes relevant in our study.} 
\label{neutrino-scatter-plot-1}
\end{figure}

The WIMP DM in our study can also be detected at the indirect detection (ID) experiments 
when DM annihilates into the SM particles. There are many experiments deployed for this
purpose like Fermi-LAT \cite{MAGIC:2016xys} and many planned experiments are 
underway like Cherenkov Telescope Array \cite{CTA:2020qlo}.
The indirect detection estimates when DM annihilation into the SM final states can be
summarised as follows,
\begin{eqnarray}
\left( \sigma v \right)_{kk} &\simeq& \frac{n_{c} v^{2}_{rel} M^2_{b} M^2_{\psi_{1}} 
\left( 1 - \frac{M^2_{b}}{M^2_{\psi_{1}}} \right)^{3/2}}{8 \pi v^{2}} \sum_{i,j = 1,2,3}A_{i} A^{*}_{j}\,,\,\,{\rm for}\,\,k=b\,,\nonumber \\  
 &\simeq& \frac{v^2_{rel} M^4_{W} \sqrt{1 - \frac{M^2_{W}}{M^2_{\psi_{1}}}} }{16 \pi v^{2}} \left(3 - \frac{4 M^2_{\psi_{1}}}{M^2_{W}} 
 + \frac{4 M^4_{\psi_{1}}}{M^4_{W}} \right) \sum_{i,j = 1,2,3}A_{i} A^{*}_{j}  
\,,\,\,{\rm for}\,\,k=W^{\pm}\,.
\label{ID-analytical}
\end{eqnarray}  
where $n_{c} = 3$ is colour charge, $v_{rel} \sim 10^{-3}$ is the DM velocity 
in the present time and $A_i$ can be expressed as,
\begin{eqnarray}
A_{i} = \frac{\alpha_{11i} U_{1i}}{\left( 4 M^2_{\psi_{1}} - M^2_{h_{i}} \right) 
+ i \Gamma_{h_{i}} M_{h_{i}} }\,.
\end{eqnarray} 
Our WIMP DM is fermionic, so we always have p-wave annihilation to
the SM final state particles which is suppressed by the square of the 
DM velocity in the present time. Therefore, our WIMP DM is hard to detect at the
present time by the indirect detection experiments and possible to explore 
in future with increased sensitivity like 
the Cherenkov Telescope Array \cite{CTA:2020qlo}.

\subsection{Collider bounds}

We have not performed a dedicated study of the present model for colliders but we can constrain our model with the other well known searches for the SM. The precise measurements of the SM Higgs observables, like the Higgs signal strength 
and Higgs invisible decay can put severe bounds on the BSM scenarios for any deviation from DM. 
So, we briefly discuss them below and take a small Higgs mixing angle to accommodate the Higgs bounds in our work.

{\bf Higgs signal strengths:}  The SM Higgs boson has been measured very precisely so any deviation from
the standard scenario will be tightly constrained. 
Bound from the Higgs signal strength measurement is one such measurement. 
We assume that Higgs produces on-shell and then it decays to some SM final
states. In these assumptions, we can factorise the SM Higgs signal strength
in two parts, one is associated with production and the second one is 
associated with its decay branching. Therefore, we can write down the SM Higgs
signal strength as,
\begin{eqnarray}
\mu &=& \mu^{Prod}_{h} \mu^{Br}_{h} \nonumber \\
&=& \frac{\sigma^{Prod}_{h_{1}}}{\sigma^{Prod}_{H_{SM}}}
\times \frac{Br_{h}}{Br_{H_{SM}}}\,.
\end{eqnarray}   
In our case, the Higgs production with the SM Higgs boson 
will be proportional to the cosine of the neutral Higgs 
mixing angle {\it i.e.} $\mu^{Prod}_{h} = \cos^{2}\theta_{12} \cos^{2}\theta_{13}$\,. On the other ratio of the branchings can be expressed as,
\begin{eqnarray}
\mu^{Br}_{h} &=& \frac{Br_{h}}{Br_{H_{SM}}} \nonumber \\
&=& \frac{\Gamma_{h_{1} \rightarrow SM}}{\Gamma_{H_{SM} \rightarrow SM}}
\times \frac{\Gamma^{tot}_{h_{1}} }{\Gamma^{tot}_{h_{1}} + \Delta \Gamma_{h_{1}} }
\end{eqnarray}
where $\Delta \Gamma_{h_{1}}$ is the extra decay mode which SM like 
Higgs may have due to the additional particles. In our work, SM like 
Higgs has similar branching with the SM Higgs, so the ratio we take is 1 in our work. In Ref. \cite{Heo:2024cif}, authors have considered LHC run 2 data from ATLAS \cite{ATLAS:2022vkf} and CMS \cite{CMS:2022dwd} detectors along with the less significant LHC run 1 \cite{ATLAS:2016neq} and 
Tevatron data \cite{Herner:2016woc} 
and combinedly found the SM Higgs signal strength at the $1\,\sigma$ 
range as,
\begin{eqnarray}
\mu^{Combine} = 1.012 \pm 0.034\,. 
\end{eqnarray} 
Therefore, at $2\sigma$ range we need to satisfy 
$\cos^{2} \theta_{12} \cos^{2} \theta_{13} \geq  0.944$ and assuming $\theta_{13}$ small, we get $\sin {\theta_{12}} \leq 0.23$\,. 

{\bf Higgs invisible decays:} We have the DM components which can be produced from the Higgs decays at the 
colliders if their masses are smaller than the half of the SM Higgs boson mass. The presence of the SM Higgs boson decays can be inferred 
from the large missing momentum along the direction of the Higgs boson momentum.
Dedicated searches for such invisible modes of the SM Higgs has put bounds on the 
Higgs invisible branching ratio ($Br_{inv}$) by $Br_{inv} < 0.16$
at $95\%$ confidence level \citep{CMS:2018yfx, CMS:2021far, CMS:2020ulv}. 

The bounds associated with the additional gauge boson $Z_{BL}$ from 
colliders, mainly LEP and LHC \cite{Carena:2004xs, Cacciapaglia:2006pk, CMS:2012umo, ATLAS:2014pcp}, 
are unimportant in our study, because we have 
considered the feeble regime of the additional gauge coupling.

\subsection{BBN bounds}

As discussed before and will be discussed in detail later,
the gauge coupling $g_{BL}$, we have considered in the feeble regime,
therefore, the additional $U(1)_{B-L}$ gauge boson, $Z_{BL}$, has the possibility
to decay after BBN which is roughly 1 sec. This decay will hamper the 
successful prediction of BBN by injecting more electromagnetic energy 
\cite{Kawasaki:2017bqm} and
hinder the nucleation process.
We have estimated that for a most liberal case 
$M_{Z_{BL}} = 1$ GeV, we need $g_{BL} > 2\times 10^{-12}$ for making $Z_{BL}$ 
decays before BBN and as we increase the $Z_{BL}$ mass the limit gets lower.
In the concourse of our study, we have taken care of the BBN bound.

\subsection{Oblique parameters}
The oblique parameters, $S, T$ and $U$, can be obtained from 
 Ref. \cite{Grimus:2008nb}, as
\begin{eqnarray}
S &=& \frac{1 }{24 \pi} 
\biggl[   U^2_{11} \ln M^2_{h_{1}}
+  U^2_{12}  \ln M^2_{h_{2}}
+ U^2_{13} \ln M^2_{h_{3}} - \ln M^2_{h_{SM}} 
+ U^2_{11} \hat{G} \left(M^2_{h_{1}}, M^2_{Z} \right)
\nonumber \\
& +& U^2_{12} \hat{G} \left(M^2_{h_{2}}, M^2_{Z} \right)
 + U^2_{13} \hat{G} \left(M^2_{h_{3}}, M^2_{Z} \right) - \hat{G} \left(M^2_{h_{SM}}, M^2_{Z} \right) 
 \biggr]\,,\nonumber \\
T &=& \frac{1}{16 \pi^{2} M^2_{W} s^2_{w}} \biggl[ 3  U^2_{11} \biggl( F\left( M^2_{Z},M^2_{h_{1}} \right)
- F\left( M^2_{W},M^2_{h_{1}} \right) \biggr)
+ 3 U^2_{12} \biggl( F\left( M^2_{Z},M^2_{h_{2}} \right)
- F\left( M^2_{W},M^2_{h_{2}} \right) \biggr)
\nonumber \\
&+& 3 U^2_{13} \biggl( F\left( M^2_{Z},M^2_{h_{3}} \right)
- F\left( M^2_{W},M^2_{h_{3}} \right) \biggr)  
 - 3 \biggl( F\left( M^2_{Z},M^2_{h_{SM}} \right)
- F\left( M^2_{W},M^2_{h_{SM}} \right) \biggr) \biggr]\,, \nonumber \\
U &=& \frac{1}{24 \pi} \biggl[ 
U^2_{11} \biggl( \hat{G} \left( M^2_{h_{1}}, M^2_{W} \right)
- \hat{G} \left( M^2_{h_{1}}, M^2_{Z} \right) \biggr)
+ U^2_{12} \biggl( \hat{G} \left( M^2_{h_{2}}, M^2_{W} \right)
- \hat{G} \left( M^2_{h_{2}}, M^2_{Z} \right) \biggr) \nonumber \\
&+& U^2_{13} \biggl( \hat{G} \left( M^2_{h_{3}}, M^2_{W} \right)
- \hat{G} \left( M^2_{h_{3}}, M^2_{Z} \right) \biggr)
- \biggl( \hat{G} \left( M^2_{h_{SM}}, M^2_{W} \right)
- \hat{G} \left( M^2_{h_{SM}}, M^2_{Z} \right) \biggr) 
\biggr]\,,
\end{eqnarray}
where $U_{ij}\,\,(i,j=1,2,2)$ is the Higgses mixing matrix defined in 
Eq. ( \ref{U-PMNS}).
As we will see since we have considered small mixing angles among the 
scalars, demanded from the Higgs collider searches, so there are small contributions
to the oblique parameters, which are safe from the current bounds on
$ S, T, U$ \cite{CDF:2013dpa, ATLAS:2017rzl, LHCb:2021bjt, CDF:2013bqv}.

\section{Results}
\label{result}

In this section, we present the main results for the parameter space in the model by considering various constraints discussed in the previous section and identify the survived region in terms of physical masses and couplings for the extra particles as well as the predicted values for the DD and ID detection cross sections.

\subsection{Constraints}
Taking into account the perturbativity bounds and the vacuum stability conditions, we consider the following ranges of the model
parameters,
\begin{eqnarray}
&& 10^{-4} \leq \theta_{12} \leq 0.23\,, \,\, 10^{-4} \leq \theta_{13,23} \leq 0.1\,, \,\, 10^{-12} \leq g_{BL} \leq 10^{-8}\,,
\,\,\theta_{L,R} = 0,
\nonumber \\
 && 
\quad 1 \leq \left( M_{h_{2,3}} - M_{h_{1}} \right)[{\rm GeV}] \leq 10^{3},\, \,\, 1 \leq \left(M_{A}-  M_{h_{1}} \right)[{\rm GeV}] \leq 10^{3}\,,  \nonumber \\
 &&
\qquad\qquad\quad 1 \leq M_{Z_{BL}}[{\rm GeV}] \leq 10^{3}\,,\,\, 10^{-12} \leq \tan\beta \leq 10^{-8}\,.\label{quartic-coupling}
\end{eqnarray}

\begin{figure}[h!]
\centering
\includegraphics[angle=0,height=5.5cm,width=5.5cm]{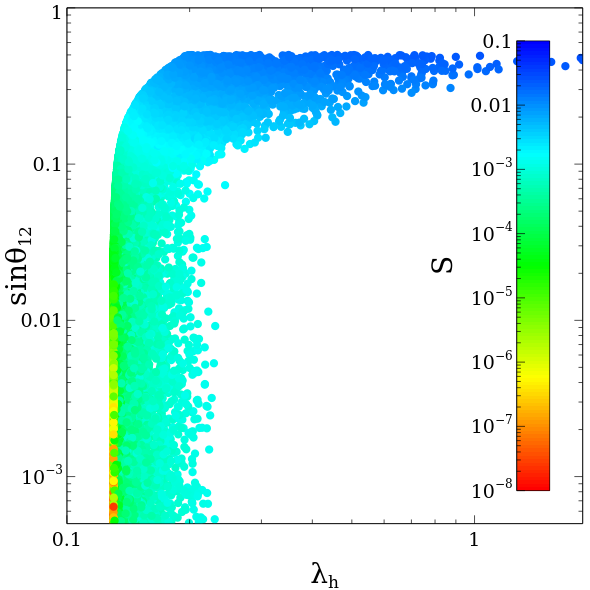}
\includegraphics[angle=0,height=5.5cm,width=5.5cm]{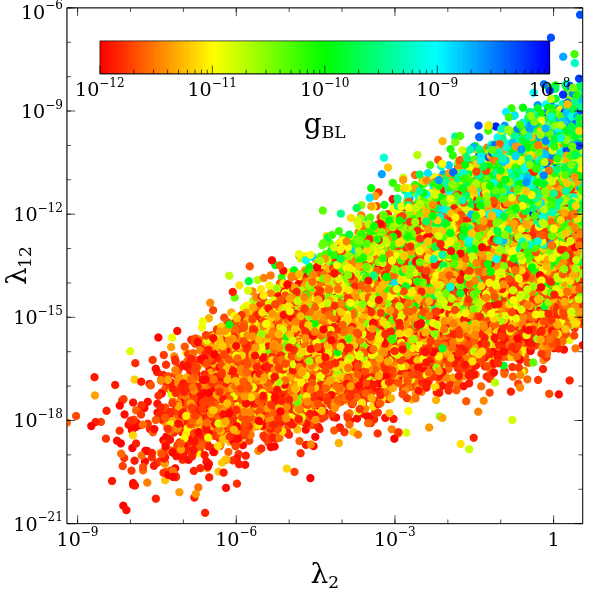}
\includegraphics[angle=0,height=5.5cm,width=5.5cm]{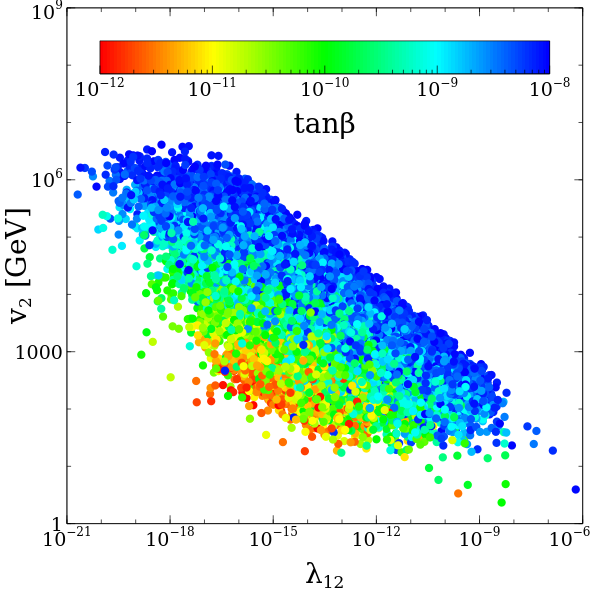}\\
\includegraphics[angle=0,height=5.5cm,width=5.5cm]{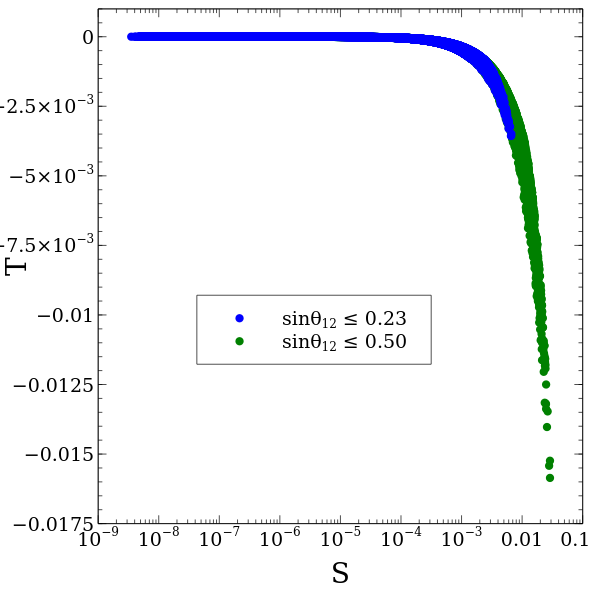}
\includegraphics[angle=0,height=5.5cm,width=5.5cm]{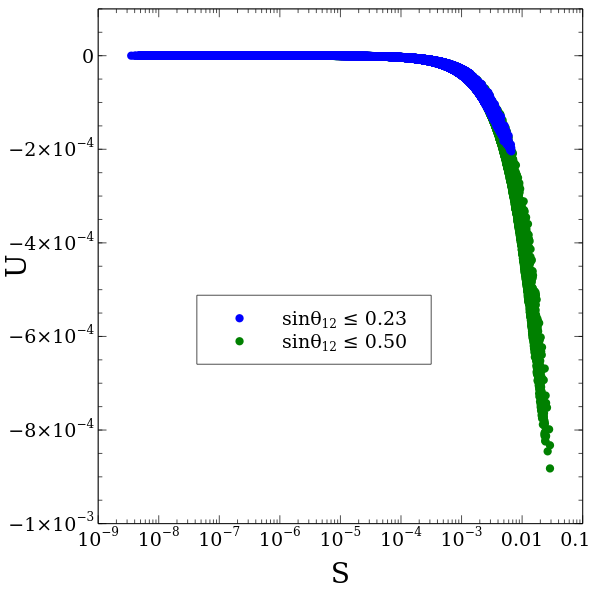}
\includegraphics[angle=0,height=5.5cm,width=5.5cm]{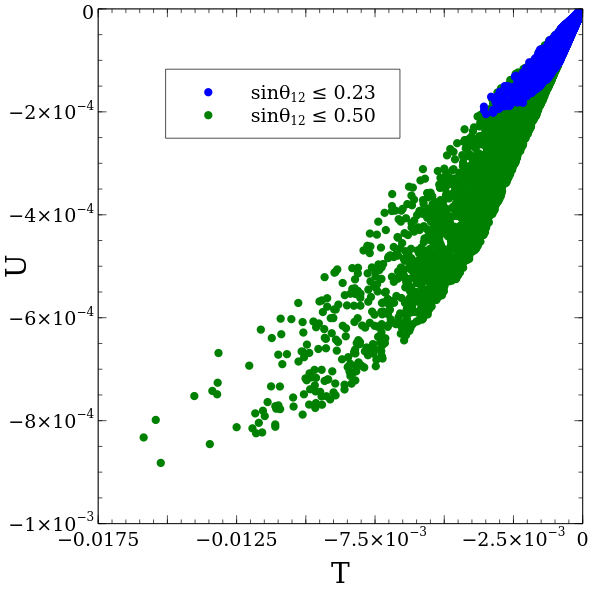}
\caption{Variation in different parameter spaces consists 
of quartic couplings, mixing angle and oblique parameters. The points have been 
obtained after satisfying the perturbativity and potential bound from the below constraints.} 
\label{quartic-couplings-allowed-range}
\end{figure}

In Fig. \ref{quartic-couplings-allowed-range}, we have shown the scattered plots among the different parameter spaces after varying the model parameters in the range shown in Eq. (\ref{quartic-coupling}).
In the top panel, we have shown the allowed region in the $\lambda_{h}-\sin\theta_{12}$, $\lambda_{2}-\lambda_{12}$, and $\lambda_{12}-v_{2}$ planes after incorporating the perturbativity and vacuum stability bounds. In the top left panel, we can see that for $\sin\theta_{12} < 0.1$, the value of $\lambda_{h}$ is independent of the increment in $\sin\theta_{12}$, and once $\sin\theta_{12}$ goes beyond 0.1, $\lambda_{h}$ starts increasing. This is because, for larger values of the mixing angle, the contribution from the BSM Higgses becomes effective. The colour bar represents the different values of the oblique parameter $S$, and it never goes beyond 0.1 for the chosen parameter range.
In the top middle panel, we have shown the plot in the $\lambda_{2}-\lambda_{12}$ plane, and the colour bar shows the different values of $g_{BL}$. We can see a correlation between the quartic couplings, and this is the effect of the $v_{2}$ VEV because both of them depend on the VEV $v_2$. From the colour bar, we can see that as the gauge coupling values increase, we observe an increase in the $\lambda_{12}$ values, which implies that the VEV $v_{1}$  decreases. However, there is no straightforward relation for $\lambda_{2}$ because the VEV $v_{2}$ is subdominant compared to $v_1$ in our study.
In the top right panel, we have shown the scatter plot in the $\lambda_{12}-v_{2}$ plane, where the colour bar shows the different values of $\tan\beta$. From the figure, we can see an anti-correlation between them, which is expected because $\lambda_{12}$ is inversely proportional to $v_{2}$. Moreover, $\tan\beta$ is linearly proportional to the VEV $v_{2}$, which can be easily seen from the colour bar.

In the lower panel of Fig. \ref{quartic-couplings-allowed-range}, we have shown the scatter plot among the oblique parameters $S$, $T$, and $U$. All the values are allowed by the existing bounds on the oblique parameters \cite{CDF:2013dpa, ATLAS:2017rzl, LHCb:2021bjt, CDF:2013bqv}, and we do not have significant deviations from the Standard Model (SM). We also observe a nice correlation among them, although they are small for most of the allowed regions. The smallness of $S$, $T$, and $U$ is mainly due to the smaller values of the mixing angles $\theta_{12}$, $\theta_{23}$, and $\theta_{13}$ which ensures the safety from the collider bound. 
We have shown the blue and red points for different upper limits of $\sin\theta_{12}$, and the allowed ranges are larger when we choose a larger upper 
limit on $\sin\theta_{12}$.

\subsection{DM results}

In the present work, we explore the alternate scenario in the context of 
$U(1)_{B-L}$ model where we have taken additional fermions instead of the 
usual right handed neutrinos 
\cite{Mohapatra:1980qe, Georgi:1981pg, Wetterich:1981bx, Lindner:2011it, Okada:2010wd} 
for the cancellation of gauge anomalies. We focus on the WIMP and FIMP combination of DM, demanding the extra gauge coupling to be in the feeble regime. The feeble coupling
also makes the additional gauge boson $Z_{BL}$ out of equilibrium and hence 
we need to consider the non-thermal nature of the distribution of $Z_{BL}$ for a treatment of the $Z_{BL}$ decay. 
However, we show later that it is sufficient to take the Maxwell-Boltzmann distribution for  $Z_{BL}$ in order to obtain the result from the $Z_{BL}$ decay.   

The distribution function for the $Z_{BL}$ can be determined from the following
Boltzmann equation,
\begin{eqnarray}
\hat{L} f_{Z_{BL}} =  \sum_{i=1,2,3}\mathcal{C}^{h_{i} \rightarrow Z_{BL} Z_{BL}}
+\sum_{B,C = A, h_{i}} \mathcal{C}^{B \rightarrow Z_{BL} C} 
+ \mathcal{C}^{Z_{BL} \rightarrow All}
\label{lioville-BE} 
\end{eqnarray}
where the Lioville's operator, $\hat{L}$, can be expressed as,
\begin{eqnarray}
\hat{L} = z H \left(1 + \frac{T g^{\prime}_{s}}{3 g_{s}} \right) \frac{\partial}{\partial z}\,.
\end{eqnarray}
In the above equation, the left-hand side corresponds to the DM evolution due to the
expansion of the Universe and the right-hand side implies the collision functions associated 
with the additional gauge boson $Z_{BL}$. The form of the collision
functions, $\mathcal{C}^{h_{i} \rightarrow Z_{BL}Z_{BL}}$, 
$\mathcal{C}^{B \rightarrow Z_{BL}C}$ and $\mathcal{C}^{Z_{BL} 
\rightarrow All}$, have been shown in the Appendix.

The Boltzmann equations for the DM candidates, $\psi_{1}$ and $\psi_{2}$,
and the next to stable particle $Z_{BL}$, can be expressed as
\begin{eqnarray}
\frac{d Y_{\psi_{1}}}{d z} &=& - \frac{S(z_{\psi_{1}}) 
\langle \sigma v \rangle_{\psi_{1}\psi_{1}}}{z_{\psi_{1}} H(z_{\psi_{1}})}
\biggl( Y^2_{\psi_{1}} - Y^{eq\,2}_{\psi_{1}} \biggr)
\nonumber \\ &-&\sum_{A=h_{i},Z_{BL}} \theta(M_{\psi_{1}} - M_{\psi_{2}} - M_{A}) 
\frac{M_{pl} z \sqrt{g_{eff}}}{0.33 M^2_{sc} g_{*,s}(z)} \biggl(
\langle \Gamma_{\psi_{1} \rightarrow \psi_{2} A} \rangle 
\left( Y^{eq}_{\psi_{1}} - Y_{\psi_{2}} Y_{A} 
\right)
\biggr)\,, \nonumber \\
\frac{d Y_{Z_{BL}}}{d z} &=& \sum_{B = h_{i}} \theta(M_{B} - 2 M_{Z_{BL}})
\frac{2 M_{pl} z \sqrt{g_{eff}}}{0.33 M^2_{sc} g_{*,s}(z)}
 \langle 
\Gamma_{B \rightarrow Z_{BL} Z_{BL} } \rangle \biggl( Y^{eq}_{B}
- Y^2_{Z_{BL}} \biggr) \biggr] 
 \nonumber \\
&+& \sum_{B,C=h_{i},A,\psi_{1}} \theta(M_{B} - M_{C} - M_{Z_{BL}}) 
\frac{M_{pl} z \sqrt{g_{eff}}}{0.33 M^2_{sc} g_{*,s}(z)} \biggl(
\langle \Gamma_{B \rightarrow C Z_{BL}} \rangle 
\left( Y^{eq}_{B} - Y_{C} Y_{Z_{BL}} 
\right)
\biggr)
\nonumber \\
 &-&
 \sum_{C = All} \theta(M_{Z_{BL}} - 2 M_{C})
  \frac{M_{pl} z \sqrt{g_{eff}}}{0.33 M^2_{sc} g_{*,s}(z)}
 \langle 
\Gamma_{Z_{BL} \rightarrow C C } \rangle_{NTH} \biggl( Y_{Z_{BL}}
- Y^2_{C} \biggr) \biggr] \,,
 \nonumber \\
\frac{d Y_{\psi_{2}}}{d z} &=& \sum_{B = h_{i}} \theta(M_{B} - 2 M_{\psi_{2}})
\frac{2 M_{pl} z \sqrt{g_{eff}}}{0.33 M^2_{sc} g_{*,s}(z)}
 \langle 
\Gamma_{B \rightarrow \psi_{2} \psi_{2} } \rangle \biggl( Y^{eq}_{B}
- Y^2_{\psi_{2}} \biggr) \biggr] 
 \nonumber \\
 &+&
  \theta(M_{Z_{BL}} - 2 M_{\psi_{2}}) \frac{2 M_{pl} z \sqrt{g_{eff}}}{0.33 M^2_{sc} g_{*,s}(z)}
 \langle 
\Gamma_{Z_{BL} \rightarrow \psi_{2} \psi_{2} } \rangle_{NTH} \biggl( Y_{Z_{BL}}
- Y^2_{\psi_{2}} \biggr) \biggr] 
 \nonumber \\
&+& \sum_{A=h_{i},Z_{BL}} \theta(M_{\psi_{1}} - M_{\psi_{2}} - M_{A}) 
\frac{M_{pl} z \sqrt{g_{eff}}}{0.33 M^2_{sc} g_{*,s}(z)} \biggl(
\langle \Gamma_{\psi_{1} \rightarrow \psi_{2} A} \rangle 
\left( Y^{eq}_{\psi_{1}} - Y_{\psi_{2}} Y_{A} 
\right)
\biggr)
\end{eqnarray}   
where 
\begin{eqnarray}
\langle \Gamma_{X\rightarrow BC} \rangle = \Gamma_{X\rightarrow BC} \frac{K_{1}(z_{X})}{K_{2}(z_{X})}\,,\,\,
 \langle \Gamma_{Z_{BL}\rightarrow BC} \rangle_{NTH}
 = M_{Z_{BL}} \Gamma_{Z_{BL}\rightarrow BC} \frac{\int \frac{f_{Z_{BL}} d^{3}p}
 {\sqrt{p^{2} + M^2_{Z_{BL}} }}}{\int f_{Z_{BL}} d^{3}p}\,.
\label{thermal-nonthermal-decay}
\end{eqnarray}

In the above Boltzmann equations, we have used MicrOMEGAs \cite{Belanger:2001fz} 
for solving the 
Boltzmann equation for the WIMP DM $\psi_{1}$. 
To use  MicrOMEGAs, we have implemented our
model in FeynRules \cite{Alloul:2013bka} and generated the 
CalcHEP \cite{Belyaev:2012qa} 
model files for the DM study. The co-moving number densities of
FIMP DM $\psi_{2}$ and the next to stable particle $Z_{BL}$ have
been computed based on our own code. We have used the decay 
processes only in the estimation of FIMP components and ignored the annihilation
contributions which are subdominant. 
contribution. We have modified the MicrOMEGAs code to implement the external codes
for FIMP DM and generated the plots. Once we have the co-moving number 
densities for $\psi_{1,2}, Z_{BL}$ ($Y_{\psi_{1,2}}, Y_{Z_{BL}}$), we can determine their relic densities by using
 the following expression \cite{Edsjo:1997bg},
\begin{eqnarray}
\Omega_{P} h^{2} = 2.755\times 10^{8}\times Y_{P} \times 
\left( \frac{M_{P}}{\rm GeV} \right)\,\,{\rm where\,\,} P = \psi_{1}, 
\psi_{2}, Z_{BL}\,.  
 \end{eqnarray}
In the next paragraph, we provide the
analytical estimation of the FIMP DM components which matches pretty well 
with the output of the Boltzmann equations with maximum $10\,\%$ deviation. 

{\bf Analytical estimates:} The FIMP DM production from the decays can be estimated by following  
Ref. \cite{Hall:2009bx}. If we consider a process $X \rightarrow P_{DM} Q$, 
then the relic density for FIMP DM we can estimate as follows,
\begin{eqnarray}
\Omega_{P_{DM}} h^{2} \simeq \frac{1.09 \times 10^{27} }{g^{3/2}_{\rho}}
\frac{M_{P_{DM}} \Gamma_{X}}{M^2_{X}}
\end{eqnarray} 
where $g_{\rho}$ is the relativistic matter {\it d.o.f} of the Universe during the 
production of DM, which is around $g_{\rho} \sim 100$ in our study. 
Therefore, we can estimate the production of $Z_{BL}$ from the
decays of the scalars and the $\psi_{2}$ production from the decays of the scalars 
and the extra gauge boson as follows,
\begin{eqnarray}
&& \Omega_{Z_{BL}} h^{2} \simeq \sum_{X = h_{1,2,3}} \frac{2.18 \times 10^{27}}{g^{3/2}_{\rho}}
\frac{M_{Z_{BL}} \Gamma_{X \rightarrow Z_{BL} Z_{BL}}}{M^2_{X}}
+ \sum_{X,Q = h_{1,2,3},A} \frac{1.09 \times 10^{27}}{g^{3/2}_{\rho}}
\frac{M_{Z_{BL}} \Gamma_{X \rightarrow Z_{BL} Q}}{M^2_{X}} \,\nonumber \\
&& \Omega_{\psi_{2}} h^{2} \simeq \sum_{X = h_{1,2,3}, A} \frac{2.18 \times 10^{27}}{g^{3/2}_{\rho}}
\frac{M_{\psi_{2}} \Gamma_{X \rightarrow \psi_{2} \psi_{2} }}{M^2_{X}}
+ 2\, Br(Z_{BL} \rightarrow \psi_{2} \psi_{2} )\,
\frac{M_{\psi_{2}}}{M_{Z_{BL}}} \left( \Omega_{Z_{BL}} h^{2} \right)\,.
\label{analytical-estimate-fimp}
\end{eqnarray}
All the decay widths needed for the analytical estimates in the above equations
have been provided in the Appendix. 
We have checked carefully that the analytical 
estimates matches perfectly well with the solution obtained after solving the full
differential equations with a maximum deviation of $10\,\%$.

\subsection{Thermal and non-thermal distributions}

\begin{figure}[h!]
\centering
\includegraphics[angle=0,height=8.5cm,width=8.5cm]{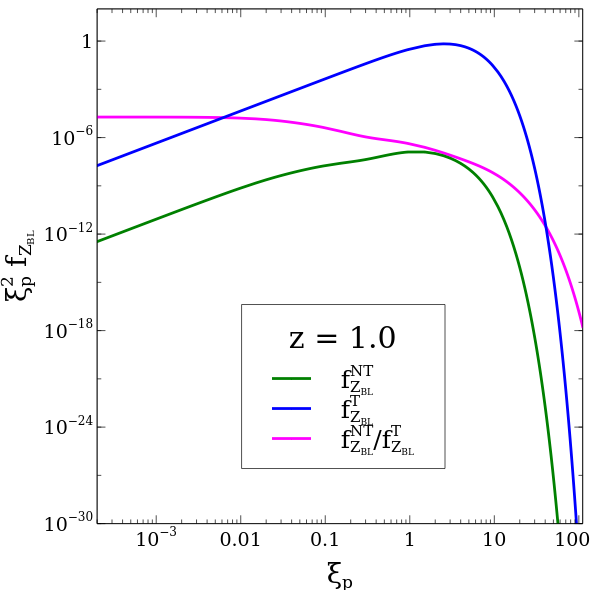}
\includegraphics[angle=0,height=8.5cm,width=8.5cm]{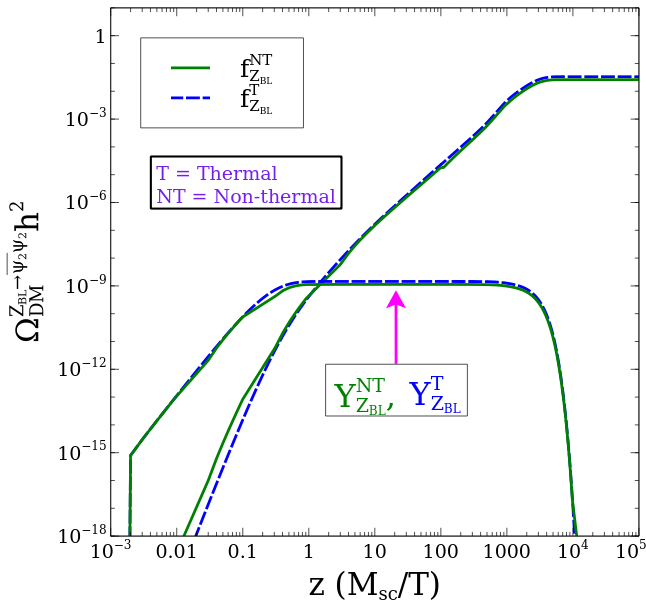}
\caption{$\psi_{2}$ DM production from the $Z_{BL}$ decay by considering thermal
and non-thermal distribution function of $Z_{BL}$. The other parameters 
have been kept fixed at $M_{h_{2,3}} = 1000$ GeV, $M_{A} = 1500$ GeV,
$\theta_{ij} = 10^{-3}$, $M_{\psi_{2}} = 1.5$ GeV, $M_{\psi_{1}} = 200$ GeV,
$g_{BL} = 10^{-11}$ and $\tan\beta = 2\times 10^{-7}$.} 
\label{thermal-non-thermal}
\end{figure}
In the left panel (LP) of Fig. \ref{thermal-non-thermal}, we have shown the 
$\xi^2_{p} f_{Z_{BL}}$ with $\xi_p$ \footnote{Defined in appendix as 
$\xi_{p} = \left( \frac{g_{s}(M_{sc}/z)}{g_{s}(M_{sc}/z_{ini})} \right)^{1/3} \frac{p M_{sc}}{z} $ where $z_{ini}$ is the initial value, $p$ is the amplitude of three momentum } for the thermal
and non-thermal distribution of $Z_{BL}$. The parameters value taken for
generating the LP and right panel (RP) have been
mentioned in the caption of Fig. \ref{thermal-non-thermal}.
The non-thermal distribution function of $Z_{BL}$ has been derived
after solving the Eq. (\ref{lioville-BE}) and for the thermal distribution
of $Z_{BL}$ we have Maxwell-Boltzmann distribution $f^{T}_{Z_{BL}} = 
e^{-\frac{E}{T}}$ where $E$ is the relativistic energy of $Z_{BL}$ 
and $T$ is the temperature. The green line shows the non-thermal distribution,
blue line shows the thermal distribution and the magenta
line shows the ratio between them. We can see the ratio 
$\frac{f^{NT}_{Z_{BL}}}{f^{T}_{Z_{BL}}}$ is nearly constant for
 $\xi_{p} < 1$ after it is exponential suppressed \footnote{
 In Ref. \cite{Decant:2021mhj} the distribution function for $z \rightarrow \infty$
 has been shown analytically which is not useful in our case because our 
 production is arounf $z \rightarrow 1$ and $z \sim 10^3$ it decays 
 to DM.}. Therefore, we can
 consider $f^{NT}_{Z_{BL}} = \kappa(\xi_p) f^{T}_{Z_{BL}}$ where
 $\kappa(\xi_p)$ is nearly constant quantity for $\xi_p < 1$.  
This proportionality
 makes sure that the non-thermal decay width of $Z_{BL}$, $\langle \Gamma_{Z_{BL}\rightarrow BC} \rangle_{NTH}$, 
as defined in Eq. (\ref{thermal-nonthermal-decay}) will be same if we use
$f^{NT}_{Z_{BL}}$ or $f^{T}_{Z_{BL}}$ because for thermal distribution
we know 
$\displaystyle\lim_{z \gg 1} \left( \frac{K_{1}(z)}{K_{2}(z)} \right) 
\rightarrow 1$, {i.e.} for high $z$ value the numerator and denominator 
converge to the same values.
In the RP of 
Fig. \ref{thermal-non-thermal}, we have established this statement.  

In the RP of Fig. \ref{thermal-non-thermal}, we have shown the comparison of the $\psi_{2}$ DM production from the decay of the $Z_{BL}$ after considering its thermal and non-thermal distribution as discussed in the LP. 
The green solid line has been plotted using the non-thermal distribution, which was derived after solving the Boltzmann equation represented by Eq. (\ref{lioville-BE}).
Basically, we have computed the distribution function of $Z_{BL}$ and, 
using it, determined the density of $Z_{BL}$. On the other hand, 
the blue dashed line has been generated using the direct decay of BSM Higgs bosons
and is independent of the $Z_{BL}$ distribution function.
So, in the solid line, we have determined the distribution function of $Z_{BL}$ and, from it, calculated its number density. In contrast, in the blue dashed line, we have directly measured the number density of $Z_{BL}$ without focusing on its distribution function. Since both originate from Higgs decay, they match each other. We observe a strong agreement between the $\psi_{2}$ DM production using the thermal and non-thermal distributions of $Z_{BL}$ which is also the finding of the LP. This can be understood by the fact that the $Z_{BL}$ decay is governed by the thermal and non-thermal
distribution have the same kind of effect because of their same kind of 
distribution with $\xi_p$ with some deviation in the magnitude. 
As mentioned, there is strong agreement\footnote{We have checked that we have
maximal 10\% deviation, which can be fixed by iteration during the computation of the differential equation, although this is time-consuming due to multiple iterations, compared to the thermal distribution where we only have a single iteration.} because, when $Z_{BL}$ decays, the non-thermal distribution of $Z_{BL}$ mimics the thermal distribution value 
\cite{Abdallah:2019svm, Biswas:2017ait}. 
Therefore, we do not see any deviation in the production of $\psi_{2}$ DM.
For careful analysis, it is therefore recommended to find the distribution function, but it is very time-consuming because we need to solve two interdependent differential equations simultaneously. In our work, 
as we obtained the two distributions provides the same output, so we continue our
analysis by assuming the thermal distribution
of $Z_{BL}$ only to consider its thermal decay width average and the 
production does not depend on the distribution function, which makes the numerical computation much faster.

\subsection{Line plots}

\begin{figure}[h!]
\centering
\includegraphics[angle=0,height=7.5cm,width=7.5cm]{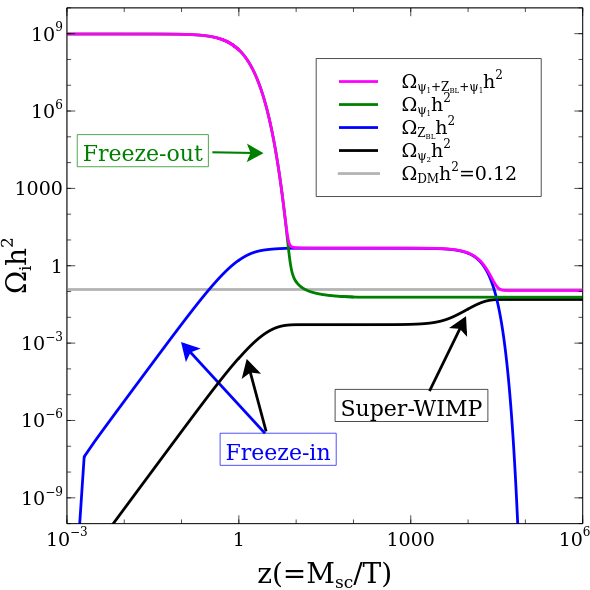}
\includegraphics[angle=0,height=7.5cm,width=7.5cm]{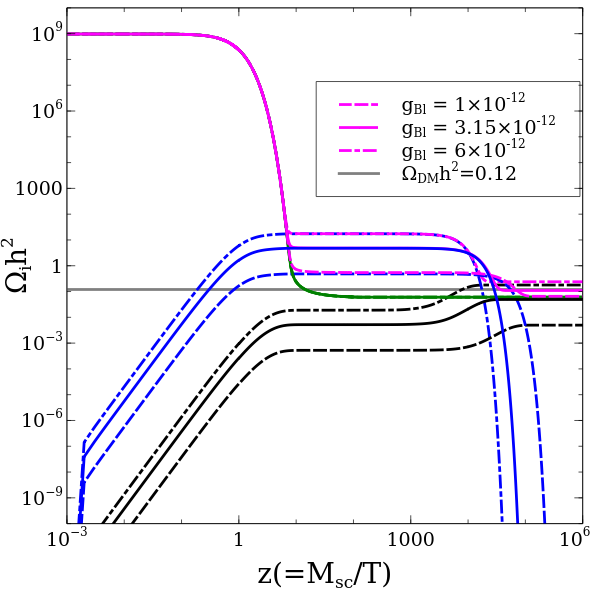}
\caption{LP shows the $\psi_{1,2}$ DM production by freeze-out, freeze-in and 
super-WIMP mechanisms whereas RP shows the response of the WIMP and 
FIMP DM production mechanisms for
 three different values gauge coupling $g_{BL}$. The parameters value 
 have been kept fixed 
 at $g_{BL} = 3.15 \times 10^{-12}$, $M_{Z_{BL}} = 24.5$ GeV, 
 $M_{\psi_{1}} = 408.4$, $M_{\psi_{2}} = 7.66$ GeV, $\tan\beta = 3.29 \times
 10^{-10}$, $M_{h_{2}} = 131.5$ GeV, $M_{h_{3}} = 346.6$ GeV,
 $M_{A} = 192.5$ GeV and Higgs mixing matrix as $U_{ii} = 1.0$ ($i=1,2,3$),
 $U_{12} \simeq 
 U_{21} = 0.011 $, $U_{13} \simeq U_{31} = 0.003$ and
 $U_{23} \simeq -U_{32} = 0.0003$. 
 The parameters have been kept fixed for the other line plots as well, 
 unless their variations are explicitly shown.} 
\label{line-plot-1}
\end{figure}

In the left panel of Fig. \ref{line-plot-1}, we have shown the $\psi_{1}$ and $\psi_{2}$ DM productions by different mechanisms, namely freeze-out 
\cite{Kolb:1985nn, Srednicki:1988ce, Gondolo:1990dk}, 
freeze-in \cite{McDonald:2001vt, Hall:2009bx}, and super-WIMP \cite{Covi:1999ty}. The green line shows the production of WIMP-type DM $\psi_{1}$, the blue line shows the $Z_{BL}$ production by the freeze-in mechanism from the decays of $h_{1,2,3}, A$, and the black line shows the freeze-in production of $\psi_{2}$ near $z \sim 1$ from the decays of $h_{1,2,3}, A$, as well as super-WIMP production near $z \sim 10^{4}$ from the decays of $Z_{BL}$. The magenta line shows the total DM production coming from both $\psi_{1}$ and $\psi_{2}$. The grey line shows the Planck 2018 \cite{Planck:2018vyg} best-fit value for the DM relic density, which coincides with the magenta line. The plot has been generated for BP1, which is also displayed in the caption of the figure.
For the subsequent line plots, we have kept the colour coding the same for the different production mechanisms and only changed the line style to indicate the change in parameter values.

In the right panel Fig. \ref{line-plot-1}, we have shown the variation in the DM production for three different values of the gauge coupling $g_{BL}$. We observe an increment in the $Z_{BL}$ and 
$\psi_{2}$ production as we increase the value of $g_{BL}$. For the variation of 
$g_{BL}$, we have kept the $Z_{BL}$ mass fixed and varied only $v_{1}$ and $v_{2}$,
 which were also kept fixed. The $Z_{BL}$ production is proportional to the square of $g_{BL}$, which we can see from the blue lines. On the other hand, $\psi_{2}$ 
 production is $\propto \frac{1}{v_{1}^2} \propto g_{BL}^2$, hence we also see 
 an increase in the $\psi_{2}$ production as we increase the value of the gauge 
 coupling $g_{BL}$. The analytical estimates for the production of $Z_{BL}$ and 
 $\psi_{2}$ are shown in Eq. (\ref{analytical-estimate-fimp}).
 We can also observe that due to the change in $g_{BL}$, the decay lifetime of 
 $Z_{BL}$ also changes, and for higher values of $g_{BL}$, the decay happens for lower values of $z$, as seen in the right panel of the figure. Moreover, we see that there is no change in the WIMP DM production, which is because the production of WIMP DM is proportional to the square of the VEV $v_{2}$, which is unchanged in this case. The total DM density is shown by the magenta points, and we observe that there is variation in the total relic density.

\begin{figure}[h!]
\centering
\includegraphics[angle=0,height=7.5cm,width=7.5cm]{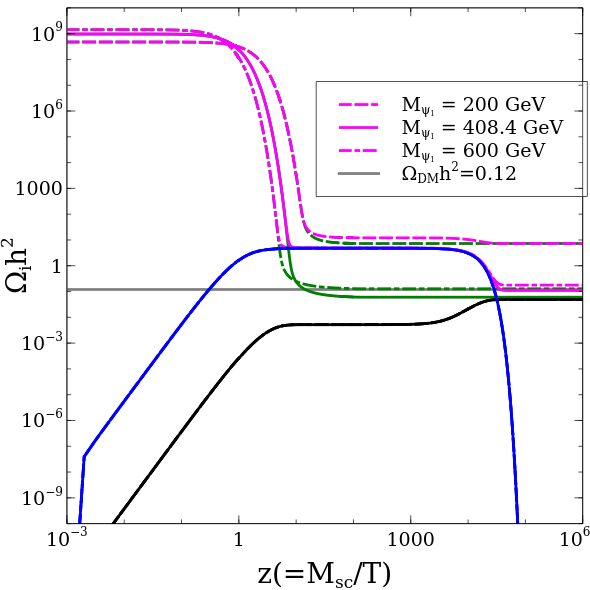}
\includegraphics[angle=0,height=7.5cm,width=7.5cm]{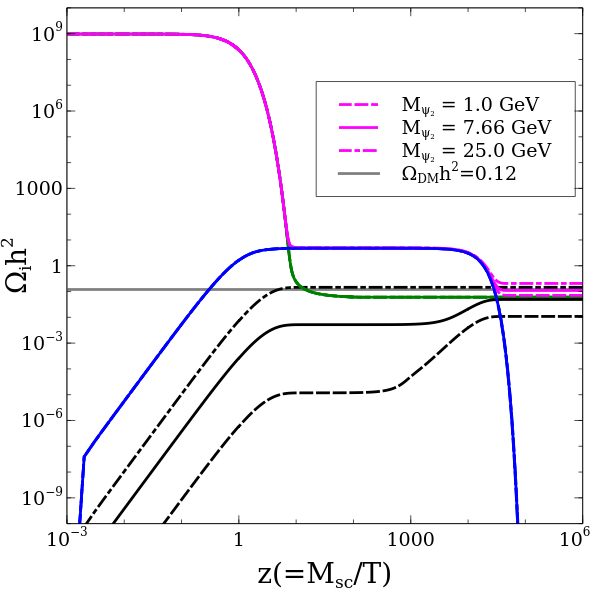}
\caption{LP and RP shows the WIMP and FIMP type DM relic density 
variation for three different values of $M_{\psi_{1}}$ and $M_{\psi_{2}}$.} 
\label{line-plot-2}
\end{figure}

In Fig. \ref{line-plot-2}, we have shown the variation of DM production for three different values of $M_{\psi_{1}}$ in the left panel and for three different values of $M_{\psi_{2}}$ in the right panel. In the left panel, we see that there is no effect on the production of the FIMP DM for the variation of $M_{\psi_{1}}$, and we observe the opposite scenario for the variation of $M_{\psi_{2}}$. In the LP, we can see that for $M_{\psi_{1}} = 200$ GeV, there is a jump in the WIMP production because $\psi_{1} \psi_{1} \rightarrow h_{3} h_{3}$ is not allowed kinematically, and $\psi_{1} \psi_{1} \rightarrow A A$ is phase space suppressed. For the other two values of $M_{\psi_{1}}$, we see an increment in the relic density for higher masses due to the reduced effect of phase space suppression, which increases the cross section times velocity and the mass. In the right panel, we can see that the $\psi_{2}$ DM relic density increases proportionally with its mass.
For the super-WIMP regime, we do not have production of $\psi_{2}$ for $M_{\psi_{2}} = 25.0$ because $Z_{BL} \rightarrow \psi_{2} \psi_{2}$ is kinematically forbidden, but it is kinematically allowed for the other two values of $M_{\psi_{2}}$, as seen in the figure. In both plots, we do not observe any effect on the production of $Z_{BL}$.

\begin{figure}[h!]
\centering
\includegraphics[angle=0,height=7.5cm,width=7.5cm]{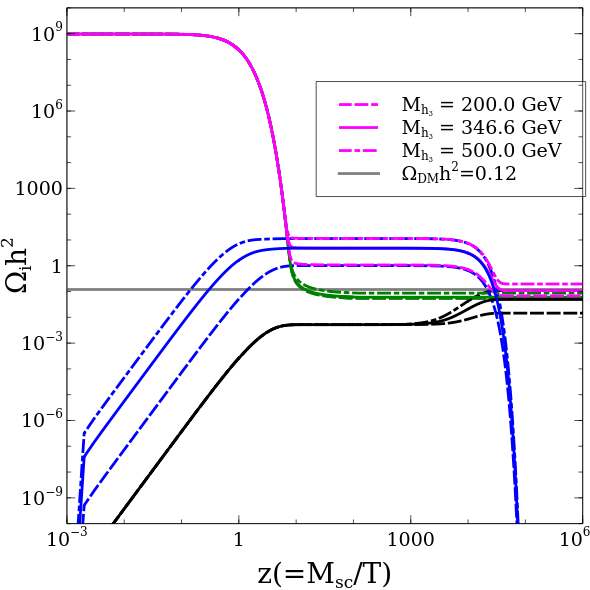}
\includegraphics[angle=0,height=7.5cm,width=7.5cm]{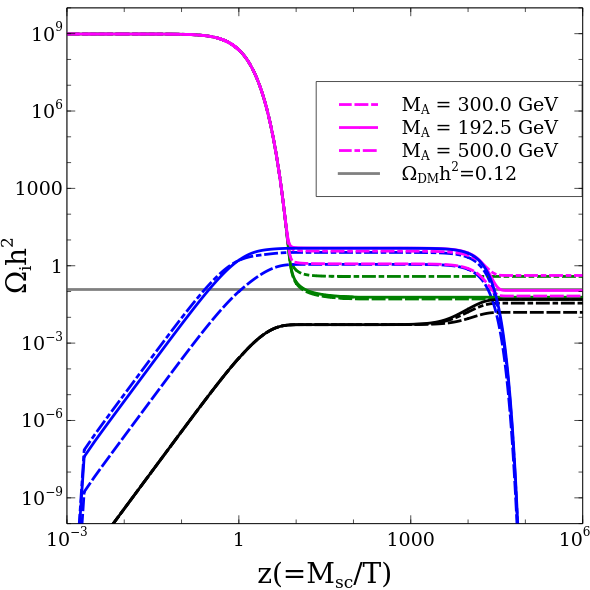}
\caption{LP (RP) shows the relic density variation for WIMP and FIMP type
DM candidates by different mechanisms for three different values of 
$M_{h_{2}}$ ($M_{A}$).} 
\label{line-plot-3}
\end{figure}

In the LP and RP of Fig. \ref{line-plot-3}, we have shown the variation in the 
DM production for three different values of $M_{h_{3}}$ and $M_{A}$, respectively. 
In both the LP and RP, we can see that there is no effect on the production 
of $\psi_{2}$ for the variation of $h_{3}$ and $A$ masses because they are 
proportional to the Higgs mixing matrix components $U_{23}$ and $\sin\beta$, which 
are small, as given in the caption of Fig. \ref{line-plot-1}. 
Therefore, the mass change does not significantly affect the production, but we have checked that the dominant production of $\psi_{2}$ comes from the $h_{2}$ decay because it is proportional to $U_{22}$.
In the case of $Z_{BL}$ production, we can see a linear growth with $M_{h_3}$, but for the RP, it is more complicated. The decay channels $A \rightarrow Z_{BL} h_{i}$ or $h_{i} \rightarrow Z_{BL} A$ contribute to $Z_{BL}$ production, depending on the masses. The dominant production mode is $h_{3} \rightarrow Z_{BL} A$ (or vice versa), depending on the masses, and other processes are subdominant due to the suppressed mixing. From $M_{A} = 192.5$ GeV to $M_{A} = 300$ GeV, there is more phase space suppression, leading to less production of $Z_{BL}$. However, when we take $M_{A} = 500$ GeV, the process $A \rightarrow h_{3} Z_{BL}$ opens up and contributes in a similar amount. This process has more phase space suppression, so it produces slightly less than for $M_{A} = 192.5$ GeV.
For the WIMP-type DM, we observe a small variation for the change in mass, but in the RP, we see a higher jump in the WIMP DM relic density for $M_{A} = 500$ GeV because the $\psi_{1} \psi_{1} \rightarrow A A$ annihilation mode is kinematically forbidden, which significantly contributes to the DM production for the chosen model parameters.

\subsection{Scattered plots}

In generating the scattered plots, we have varied the model parameters
in the following range,
\begin{eqnarray}
&& 10^{-4} \leq \theta_{ij}\,\,(i,j = 1,2,3) \leq 10^{-1}\,,
 1 \leq \left( M_{h_{2,3}} - M_{h_{1}} \right)\,\,[{\rm GeV}] \leq 10^{3}\,,
10^{-12} \leq g_{BL} \leq 10^{-8}\,,\nonumber \\
&& 1 \leq M_{Z_{BL}}\,\,[{\rm GeV}] \leq 10^{3}\,,
1 \leq \left( M_{A} - (M_{Z_{BL}} + M_{h_{1}}) \right)\,\,[{\rm GeV}]
\leq 10^{3}\,,10^{-12} \leq \tan\beta \leq 10^{-6}\,,\nonumber \\
&& 1 \leq M_{\psi_{1}} \,\,[{\rm GeV}] \leq 10^{3}\,,
1 \leq M_{\psi_{2}} \,\,[{\rm GeV}] \leq 10^{3}\,,
\theta_{L} = 0\,.
\end{eqnarray}
We have chosen the CP-odd Higgs mass 
$M_A$ within the above range so that it always has 
two-body decay channel open, avoiding suppression from a three-body decay, 
which could conflict with the BBN bound.
We have demanded that the total DM density for WIMP and FIMP DM candidates 
from all the mechanisms satisfy the following Planck $5\sigma$
range \cite{Planck:2018vyg},
\begin{eqnarray}
0.1116 \leq \left( \Omega_{\psi_{1}} + \Omega_{\psi_{2}} \right) h^{2} \leq 
0.1284\,.
\end{eqnarray}

\begin{figure}[h!]
\centering
\includegraphics[angle=0,height=7.5cm,width=7.5cm]{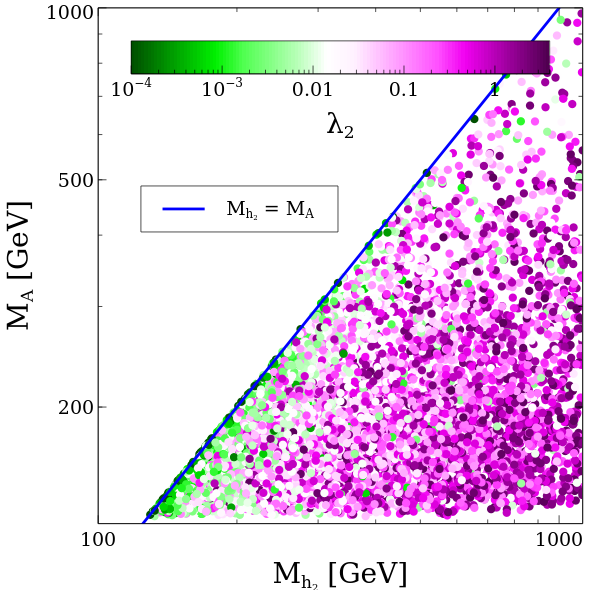}
\includegraphics[angle=0,height=7.5cm,width=7.5cm]{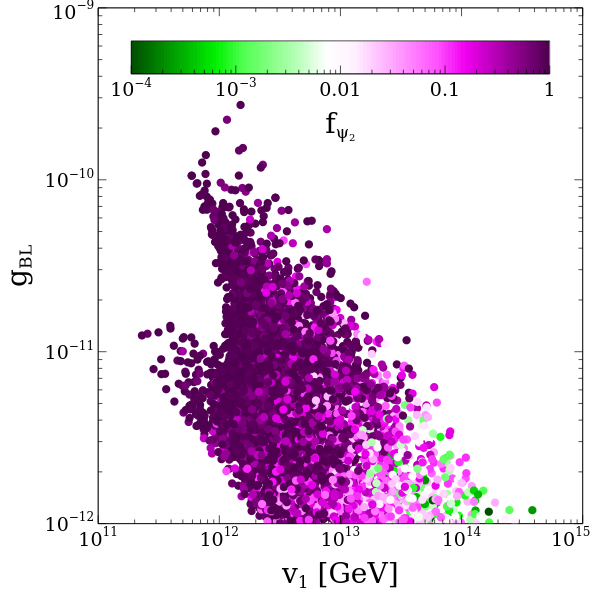}
\caption{LP and RP shows the scatter plots in the $M_{h_{2}}-M_{A}$
and $v_{1}-g_{BL}$ planes after satisfying the DM relic density and 
other constraints mentioned in section \ref{constraints}. The colour bar in the
LP shows the different values quartic coupling $\lambda_{2}$ whereas the 
RP shows the fraction of the WIMP DM contribution.} 
\label{scatter-plot-1}
\end{figure}

In the LP and RP of Fig. \ref{scatter-plot-1}, we have shown the scatter plots in the $M_{h_{2}}-M_{A}$ and $v_{1} - g_{BL}$ planes, respectively. All the points satisfy the bounds listed in section \ref{constraints} along with the DM relic density.
In the LP, we can see a sharp correlation between the values of $M_{h_{2}}$ and $M_{A}$, and the colour bar shows the different values of the quartic coupling $\lambda_{2}$. For the potential to be bounded, we need $\lambda_{2} > 0$, and the points above the blue line will provide negative values, so those are excluded. Moreover, as we go towards the x-axis, we get higher values of $\lambda_{2}$, which can be understood from
 Eq. (\ref{quartic-coupling-expression}).
In the RP, we can see scatter plots in the $v_{1}-g_{BL}$ plane, where the colour variation shows the fraction of FIMP DM $f_{\psi_{2}}$. We can see from the figure that as we go towards higher values of $g_{BL}$, we have more production of FIMP-type DM $\psi_{2}$, which can be understood from the analytical estimates in 
Eq. (\ref{analytical-estimate-fimp}). 
Moreover, as we move along the x-axis, we see a smaller fraction of $\psi_2$ DM because the production of $\psi_2$ from the Higgses is inversely proportional to the square of the VEV $v_1$. We also see an anti-correlation between $v_{1}$ and $g_{BL}$, which is expected because they appear as a product in the $Z_{BL}$ mass.

\begin{figure}[h!]
\centering
\includegraphics[angle=0,height=7.5cm,width=7.5cm]{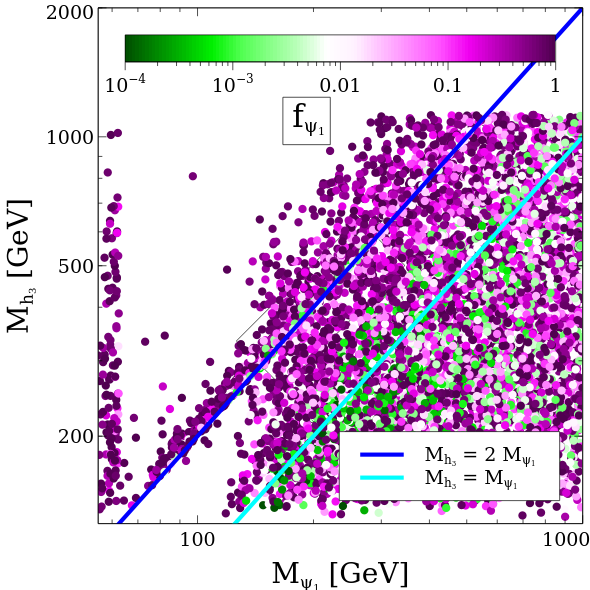}
\includegraphics[angle=0,height=7.5cm,width=7.5cm]{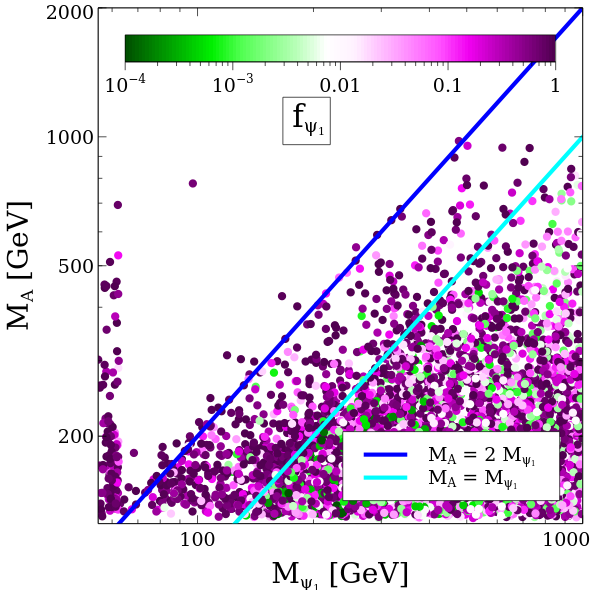}
\caption{LP shows the allowed region in the $M_{\psi_{1}} - M_{h_{3}}$ plane 
after satisfying all the relevant bounds. In the RP, we have shown the 
variation in the $M_{\psi_{1}}-M_{A}$ plane. In both plots, 
colour variation shows the fraction of the WIMP DM component.} 
\label{scatter-plot-2}
\end{figure}

In the LP of Fig. \ref{scatter-plot-2}, we have shown a scatter plot in the $M_{\psi_{1}} - M_{h_{3}}$ plane, where the colour bar represents the different values of the fraction of WIMP DM $\psi_{1}$ in the total DM density. Near $M_{\psi_{1}} \sim 62$ GeV, we see a vertical line of points, which are due to the Higgs resonance region. The points are mostly magenta because the cross-section is suppressed by the Higgs mixing angle $U_{31}$, and hence the dominating contribution to the DM density is from this suppression, rather than the strong effect of resonance. The blue line shows the resonance region between $\psi_{1}$ and $h_3$. We can see that all the points around the blue line are magenta points and there are no green points. This is because $h_3$ has a larger decay width, so the resonance effect is mild, which results in the dominating contribution to the DM density. The cyan line represents when $M_{\psi_{1}} \sim M_{h_{3}}$. Below the cyan line, the $\bar \psi_{1} \psi_{1} \rightarrow h_{3} h_{3}, AA$ annihilation modes are open and contribute dominantly, and we can see green points around the cyan line. The region between the blue and cyan lines is mainly due to the DM annihilation to $W^{+}W^{-}, ZZ, h_{2}h_{2}, h_{1}h_{1}$. The main advantage of the region below is that we can achieve DM density for a larger region of the parameter space, which will be pretty safe from direct detection due to the suppression by the mixing angle, which does not have any effect on DM production in this part of the parameter space. In the RP, we can see the resonance points near half of the SM Higgs mass. We do not see any points beyond the blue line because those points are not allowed by the perturbativity bound coming from the $\lambda_{2}$ quartic coupling, as discussed before. The cyan line represents $M_{\psi_{1}} \sim M_{A}$, and below this line, we have the $\bar \psi_1 \psi_1 \rightarrow AA$ annihilation mode open. These two plots are important from the present-day severe DD bound because we can choose $M_{\psi_{1}} > M_{h_{3}}, M_{A}$ and satisfy the DM relic density very easily, but at the same time, we can evade the direct detection bound by choosing a smaller mixing angle, which has negligible effect on the DM density.

\begin{figure}[h!]
\centering
\includegraphics[angle=0,height=7.5cm,width=7.5cm]{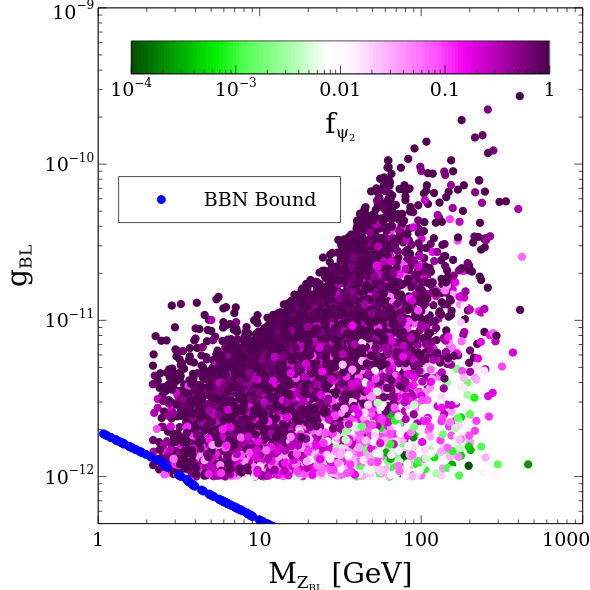}
\includegraphics[angle=0,height=7.5cm,width=7.5cm]{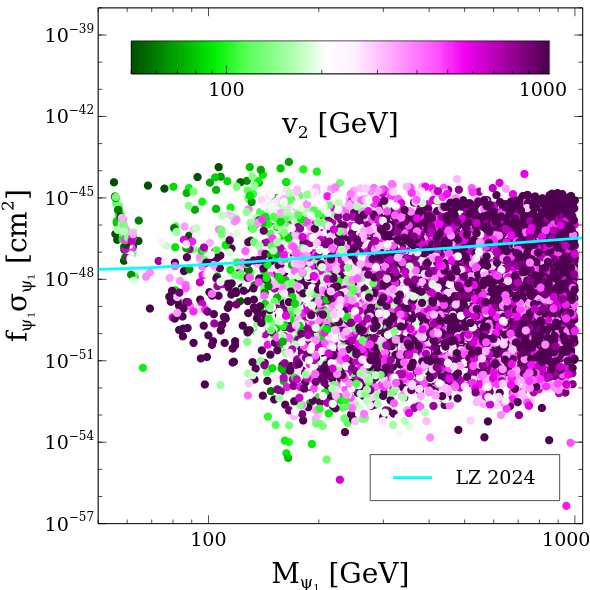}
\caption{LP shows the scatter plot in the $M_{Z_{BL}} - g_{BL}$ plane and the
colour variation represents the fraction of the FIMP DM component. In the RP,
we have shown variation in the WIMP DM mass with the SI cross-section 
where the colour variation shows the different values of VEV $v_{2}$.} 
\label{scatter-plot-3}
\end{figure}

In the LP of Fig. \ref{scatter-plot-3}, the scatter plot in the $M_{\zmt}-g_{BL}$ plane is shown, where the colour bar displays the different fractions of FIMP DM $\psi_2$ contributing to the total DM density. We see a linear relation between $M_{\zmt}$ and $g_{BL}$, which is expected from the gauge boson mass dependence on $g_{BL}$, and the width in the correlation arises due to the different values of the VEVs $v_{1,2}$. With the increase of $g_{BL}$, we see an increment in the $\psi_2$ DM fraction. The effect of the $\zmt$ mass on the $\psi_2$ production is not very prominent. The blue line represents the BBN bound, and below that line, $Z_{BL}$ decays after BBN and can alter the successful prediction of BBN.
In the RP, we have shown the scatter plot in the $M_{\psi_{1}} - f_{\psi_{1}} \sigma_{\psi_{1}}$ plane, and the colour bar shows the different values of $v_{2}$. 
The analytical expression for DD is given in Eq. (\ref{DD-expression}). 
Some parts of the parameter space are already ruled out by the LUX-ZEPLIN 2024 data \cite{LZCollaboration:2024lux},
and the rest will be explored in the near future. As mentioned before, the present model is difficult to rule out by direct detection fully because we can choose a smaller mixing angle between the DM and visible sector, which does not affect the DM relic density. Therefore, by choosing a smaller mixing angle, we can evade the bounds without altering the DM density.

\begin{figure}[h!]
\centering
\includegraphics[angle=0,height=7.5cm,width=7.5cm]{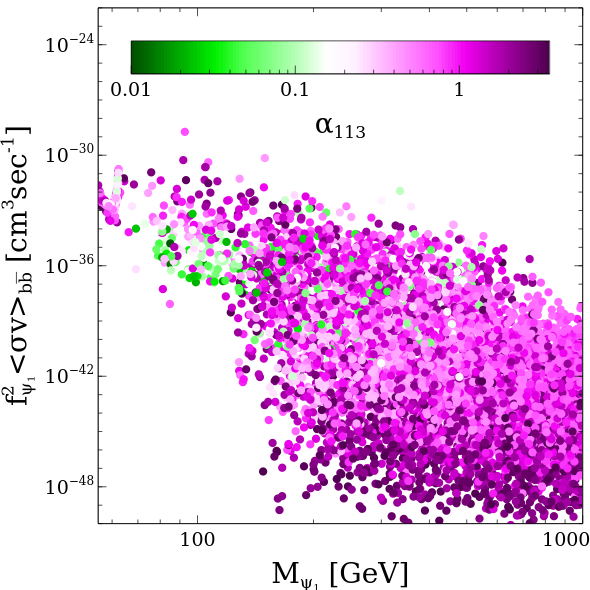}
\includegraphics[angle=0,height=7.5cm,width=7.5cm]{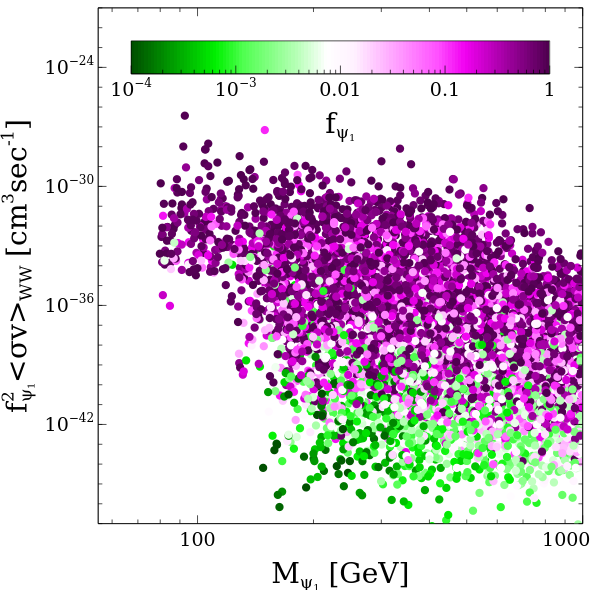}
\caption{LP  and RP show the variation of the WIMP DM mass $M_{\psi_{1}}$ with
the DM annihilation cross section to $b \bar{b}$ and $W^{+}W^{-}$ final states.
The colour variation in the LP shows the coupling strength $\alpha_{113}$ and the
RP shows the fraction of WIMP DM contribution to the total DM density.} 
\label{scatter-plot-4}
\end{figure}

In Fig. \ref{scatter-plot-4}, we have shown scatter plots in the $M_{\psi_{1}} - f^2_{\psi_{1}} \langle \sigma v \rangle_{b\bar{b}}$ and $M_{\psi_{1}} - f^2_{\psi_{1}} \langle \sigma v \rangle_{WW}$ planes. The analytical expressions for indirect detection are shown in Eq. (\ref{ID-analytical}).
 We can see that our DM is fermionic, so from the initial state wave function contraction and averaging over the initial spin, we get an amplitude $\propto (s - 4 M^2_{\psi_{1}})$, which is $\propto v^{2}_{\text{rel}}$ in the non-relativistic limit.
At present, the DM velocity is taken as $v_{\text{rel}} \sim 10^{-3}$, which makes the indirect detection cross section very small compared to the present-day bound, making it difficult to detect in the near future. In the LP, we have shown the colour variation for the Yukawa coupling value $\alpha_{113}$, and we can see that a lower value of it contributes to higher values of the ID cross section because we have multiplied by $f^2_{\psi_{1}}$, which is inversely proportional to the fourth power of the Yukawa coupling $\alpha_{113}$.
In the RP, we have shown in the colour bar the fraction of $f_{\psi_{1}}$, 
and we also see a linear relation with the ID cross-section. Therefore, our present model is very much safe from both direct and indirect detection experiments and is very timely in the context of the recent LUX-ZEPLIN data, which has ruled out a large part of the parameter space.

\section{Conclusions}
\label{conclusion}
In the present work, we have considered the extension of the SM with the $U(1)_{B-L}$ gauge symmetry, but introduced four chiral fermions for gauge anomaly cancellation, instead of three right-handed fermions. Moreover, the scalar sector has been extended by two singlet scalars, whose VEVs provide masses to the extra fermions and the  $U(1)_{B-L}$ gauge boson. From the four chiral fermions, we can compose two Dirac fermions, which are candidates for WIMP and FIMP DM in our work. We can take the mixing among the chiral components of the two Dirac fields to zero by introducing two independent $Z_2$ symmetries for the extra fermions or considering the associated Yukawa 
couplings to zero. 
We can achieve many different kinds of possibilities for DM types and productions, but we focused on the WIMP and FIMP DM combination, which could survive even after more severe bounds in the future. We have taken the hierarchical VEVs of the singlet scalars, $v_{1} \gg v_{2}$, to realize the WIMP and FIMP combination at the renormalizable level.

In our model, we considered the mass of the extra gauge boson $Z_{BL}$ in the TeV scale, so we assumed the associated extra gauge coupling in the feeble regime. This makes the extra gauge boson out of equilibrium, so we have computed its non-thermal distribution function for that. However, we found that our analysis does not change even if we consider the thermal distribution for the gauge boson, because the ratio of non-thermal and thermal distributions of the gauge boson remains fixed for the distribution function with $\xi_{p} < 1$, and the $\zmt$ number density falls exponentially for larger values of $\xi_{p}$. The  $Z_{BL}$ gauge boson can be produced dominantly from the decays of SM Higgs and singlet scalars in the early Universe, while the annihilation contributions to the abundance of  the $Z_{BL}$ gauge boson can be neglected. We showed that the FIMP DM $\psi_2$ can be produced from the late decay of $Z_{BL}$ and scalar fields in our model, while the BBN bound is satisfied. The annihilation contributions to the production of FIMP DM can be also neglected.

In the case of WIMP DM, we have focused on the regions of the parameter space where $\psi_1 \psi_1 \rightarrow h_{i} h_{i}, AA$ annihilation modes are active on top of the Higgs resonance region. Since DM mainly annihilates into $AA$ and $h_{3} h_{3}$, we can freely choose the mixing angles that connect the visible and dark sectors. Therefore, a large portion of the parameter space for WIMP DM is still unexplored by existing direct detection experiments but can be probed in future experiments. 
Moreover, as the WIMP DM is fermionic in nature, the annihilation cross section for WIMP into the SM sector is $p$-wave suppressed, so it is challenging to probe the WIMP DM by indirect detection experiments. We have shown the allowed regions in various planes for the parameter space after implementing all the relevant constraints.

Moreover, we also showed the possibility of generating neutrino masses by dimension-6 and dimension-7 operators in the effective theory with the $U(1)_{B-L}$ symmetry. These higher-dimensional operators can be realized when the model is extended with three singlet fermions, which are suitably charged under  $U(1)_{B-L}$ for gauge anomaly conditions and have the Yukawa interactions with the lepton doublets either by dimension-4 or dimension-5 terms. It would be worthwhile to explore our model further such as alternative production mechanisms for DM and detection prospects at the present and future colliders.   

\section*{Acknowledgements}
The research is supported by Brain Pool program funded by the Ministry of Science and ICT through the National Research Foundation of Korea\,(RS-2024-00407977) and Basic Science Research Program through the National Research Foundation of Korea (NRF) funded by the Ministry of Education, Science and Technology (NRF-2022R1A2C2003567).
For the numerical analysis, we have used the Scientific Compute Cluster at GWDG, the joint data center of Max Planck Society for the Advancement of Science (MPG) and University of G\"{o}ttingen.
\appendix
\section*{Appendix: Analytical expressions for decay widths and collision terms}
\label{App:AppendixA}
If we consider a generic process, $\chi (\tilde{p}) \rightarrow a(\tilde{p_1})\,
b(\tilde{p_2})$ (where $\tilde{p} = (E_{p},\bar{p})$),
then the collision term can be written in the
following form \cite{Kolb:1990vq, Gondolo:1990dk},
\begin{eqnarray}
\mathcal{C}[f_{\chi}(p)]&=&\dfrac{1}{2\,E_p}\int
\dfrac{g_a\,d^{3} p1}{(2\pi)^{3} \,2E_{p1}}
\dfrac{g_b\,d^{3} p2}{(2\pi)^{3} \,2E_{p2}}
(2\pi)^4\,\delta^4(\tilde{p}-\tilde{p_1}-\tilde{p_2})
\times\overline{\,\lvert \mathcal{M}\rvert^2}\nonumber \\
&&~~~~~~~~~~~~~~\times\,[f_a\,f_b\,\left(1 \pm f_\chi\right)
-f_\chi\left(1 \pm f_a\right)\left(1 \pm f_b\right)]\,.
\label{colision1}
\end{eqnarray}
Now the full expressions of the collision terms in
Eq. (\ref{lioville-BE}) are as
follows \cite{Konig:2016dzg, Biswas:2016iyh},
\begin{itemize}
\item $\mathcal{C}^{\zmt \rightarrow all}$:
\,\,Collision term for the extra gauge boson $\zmt$ total decay takes the
the following form, 
\begin{eqnarray}
\mathcal{C}^{\zmt \rightarrow all}&=&-f_{\zmt}(\xi_p)\times
\Gamma_{\zmt \rightarrow all}\times
\dfrac{z_{\zmt}}{\sqrt{\xi_p^2\,\mathcal{B}(z)^2+z_{\zmt}^2}}\,.
\end{eqnarray}
where $z_{\zmt} = \frac{M_{\zmt}}{T}$, $\mathcal{B}(z) = \left(\frac{g_{s}(M_{sc}/z)}{g_{s}(M_{sc}/z_{ini})}\right)^{1/3}$, $\xi_{p} = \mathcal{B}^{-1}(z) 
 \frac{p}{T}$, $\Gamma_{\zmt \rightarrow all} = \Gamma_{\zmt \rightarrow f \bar{f}}$
+ $\Gamma_{\zmt \rightarrow \psi_{i} \psi_{j}}$ and the expression for the each decay terms
are as follows,
\begin{eqnarray}
\Gamma_{\zmt \rightarrow f \bar{f}} &=& n_{c}
\frac{M_{\zmt} \,Q^2_{f}g_{BL}^{2}}{12\,\pi}
\left( 1 + \frac{2 M_{f}^{2}}{M_{\zmt}^{2}} \right)
\sqrt{1 - \frac{4 M_{f}^{2}}{M_{\zmt}^{2}}}, \nn \\ 
\Gamma_{\zmt \rightarrow \psi_{i} \psi_{j}} &=& 
\frac{M_{\zmt}}{12 \pi} \sqrt{\biggl[ 1 - \left( \frac{M_{\psi_{1}} + M_{\psi_{2}}}
{M_{Z_{BL}}} \right)^{2} \biggr] \biggl[ 1 - \left( \frac{M_{\psi_{1}} - M_{\psi_{2}}}
{M_{Z_{BL}}} \right)^{2} \biggr]} \nonumber \\
&\times& \biggl[ \left( g^{V}_{\zmt \psi_{i} \psi_{j}} \right)^{2} 
\biggl(1 - \left(\frac{M_{\psi_{1}} - M_{\psi_{2}} }{M_{\zmt}} \right)^{2}  \biggr)
\times \biggl(1 - \frac{1}{2}\left(\frac{M_{\psi_{1}} + M_{\psi_{2}} }{M_{\zmt}} \right)^{2}  \biggr)
\nonumber \\
&+&  \left( g^{A}_{\zmt \psi_{i} \psi_{j}} \right)^{2} 
\biggl(1 - \left(\frac{M_{\psi_{1}} + M_{\psi_{2}} }{M_{\zmt}} \right)^{2}  \biggr)
\times \biggl(1 + \frac{1}{2}\left(\frac{M_{\psi_{1}} - M_{\psi_{2}} }{M_{\zmt}} \right)^{2}  \biggr)
 \biggr]
\label{dkz}
\end{eqnarray} 
where $f$ is all the SM fermions and $Q_f$ is associated charge under $U(1)_{B-L}$. 
The coupling of $Z_{BL}$ with the BSM fermions ($\psi_{1,2}$) takes the following form,
\begin{eqnarray}
&& g^{V}_{Z_{BL} \psi_{1} \psi_{1}} = \frac{g_{BL}}{6} \left( 1 - 3 \cos^{2}\theta_{L} \right), \quad g^{A}_{Z_{BL} \psi_{1} \psi_{1}} = \frac{g_{BL}}{6} \left( 3 + 3 \cos^{2}\theta_{L} \right)\,, \nonumber \\
&& g^{V}_{Z_{BL} \psi_{1} \psi_{2}} = -\frac{g_{BL}}{3} \sin^{2}\theta_{L}, \quad g^{A}_{Z_{BL} \psi_{1} \psi_{2}} = \frac{g_{BL}}{3} \sin^{2}\theta_{L}\,, \nonumber \\
&& g^{V}_{Z_{BL} \psi_{2} \psi_{2}} = \frac{g_{BL}}{6} \left( 1 - 3 \sin^{2}\theta_{L} \right),\quad g^{A}_{Z_{BL} \psi_{1} \psi_{1}} = \frac{g_{BL}}{6} \left( 3 + 3 \sin^{2}\theta_{L} \right)\,.
\label{zmt_coup}
\end{eqnarray}

\item $\mathcal{C}^{X \rightarrow Z_{BL} Y}$:\,\,
The collision term for the production of the extra gauge boson $\zmt$ via the decay of the BSM scalars $h_i, A$ is expressed as follows,
\begin{eqnarray}
\mathcal{C}^{X \rightarrow Z_{BL}Y} &=&
\dfrac{z}{16 \pi M_{sc}}\dfrac{\mathcal{B}^{-1}(z)}
{\xi_p \sqrt{\xi_p^2\mathcal{B}(z)^2+
\left(\dfrac{M_{\zmt}\,z}{M_{sc}}\right)^2}} 
\frac{\left|M \right|^{2}_{X\rightarrow Z_{BL} Y}}{g_{Z_{BL}}}
 \nonumber \\ 
&&\times \left(e^{-\sqrt{\left(\xi_{k}^{\rm min}\right)^2
\mathcal{B}(z)^2+\left(\frac{M_{h_2}\,z}{M_{sc}}\right)^2}}
\,-\,e^{-\sqrt{\left(\xi_{k}^{\rm max}\right)^2
\mathcal{B}(z)^2+\left(\frac{M_{h_2} z}{M_{sc}}\right)^2}}
\right) \,.
\label{ch2zblzbl-final}
\end{eqnarray}

where $g_{Z_{BL}} = 3$, the squared decay amplitude, $|M|^{2}_{X\rightarrow \zmt Y}$, 
is without any average over initial and final states, and 
the symmetric factor automatically cancels when we 
multiply by the same final states, $Z_{BL} Z_{BL}$. The other relevant quantites 
for producing the DM density are defined below,

\begin{eqnarray}
|M|^{2}_{h_{i}\rightarrow \zmt\zmt} &=&g_{h_i\zmt\zmt}^2
\left(2+\dfrac{(M_{h_2}^2-2M_{\zmt}^2)^2}{4M_{\zmt}^4}\right)\,,\nonumber \\
|M|^{2}_{P\rightarrow \zmt Q} &=&g_{P\zmt Q}^2
\left( M^2_{P} - (M_{Q}-M_{\zmt})^{2} \right)
\left( M^2_{P} - (M_{Q}+M_{\zmt})^{2} \right)
\,\,{\rm with\,\,}P,Q = h_{i}, A, \nonumber \\ 
g_{h_{i}\zmt \zmt} &=& 2 g^{2}_{BL} \left( U_{2i} v_{2} + 4 U_{3i} v_{1} \right),\,\,\, 
g_{A\zmt h_{i}} = g_{BL} \left( 2 U_{3i} \cos\beta + U_{2i} \sin\beta \right),
\,\,\,\,i=1,2,3 ,
\nn \\
\xi_k^{\rm min} (\xi_p,z)&=&\dfrac{M_{sc}}{2\,\mathcal{B}(z)\,z\,M_{\zmt}}
\left| \,\eta (\xi_p,z)-\dfrac{\mathcal{B}(z)
\times M_{h_2}^2}{M_{\zmt} \times M_{sc}}\,\xi_p\,z
\right| \,,\nn \\
\xi_k^{\rm max} (\xi_p,z)&=&\dfrac{M_{sc}}{2\,\mathcal{B}(z)\,z\,M_{\zmt}}
\bigg( \,\eta (\xi_p,z)+\dfrac{\mathcal{B}(z)
\times M_{h_2}^2}{M_{\zmt}\times M_{sc}}
\,\xi_p\,z \,\bigg)\,,\nn \\
\eta(\xi_p,z)&=& \left(\frac{M_{h_2}\,z}{M_{sc}}\right)
\,\sqrt{\biggl[ \left( \frac{M_{X}}{M_{\zmt}} +1 \right)^{2} - 
\left( \frac{M_{Y}}{M_{\zmt}} \right)^{2} \biggr]
\biggl[ \left(1- \frac{M_{\zmt}}{M_{X}} \right)^{2} - 
\left( \frac{M_{\zmt}}{M_{X}} \right)^{2} \biggr]
}\,\,
\nonumber \\
&\times &
\sqrt{\xi_p^2\,\mathcal{B}(z)^2+
\left(\frac{M_{\zmt}\,z}{M_{sc}}\right)^2}\,.
\label{XZBLY}
\end{eqnarray}
Here, the above expressions match with those in Ref. \cite{Biswas:2017ait} 
for the same final state particles.

\item The other relevant decay widths, which have been used in our study, are as
follows.
\begin{itemize}
\item Decay width for $h_{i} \rightarrow \zmt \zmt $ can be expressed as follows,
\begin{eqnarray}
\Gamma_{h_{i} \rightarrow \zmt \zmt} = \frac{M^{3}_{h_{i}} g^2_{h_{i} \zmt \zmt} }
{128 \pi M^4_{\zmt}} \sqrt{1 - \frac{4 M^2_{\zmt}}{M^2_{h_{i}}}}
\left( 1 - \frac{4 M^2_{\zmt}}{M^2_{h_{i}}} + \frac{12 M^4_{\zmt}}{M^4_{h_{i}}} \right)
\end{eqnarray}
where the vertex factors $g_{h_{i}\zmt\zmt}$ are given in Eq. (\ref{XZBLY})\,.

\item The decay width for $X \rightarrow \zmt Y$, where $X,Y = h_{i}, A$ depending on
masses, can be expressed as
\begin{eqnarray}
\Gamma_{X \rightarrow \zmt Y} = \frac{g^2_{X\zmt Y} M^3_{X}}{16 \pi M^2_{\zmt}}
\biggl[ 1 - \left(\frac{M_{\zmt} + M_{Y}}{M_{X}} \right)^{2} \biggr]^{3/2}
\biggl[ 1 - \left(\frac{M_{\zmt} - M_{Y}}{M_{X}} \right)^{2} \biggr]^{3/2}
\end{eqnarray}
where the coupling $g_{X\zmt Y}$ is defined in Eq. (\ref{XZBLY}).
\item The decay width of $h_i$ to BSM fermions can be expressed as,
\begin{eqnarray}
\Gamma_{h_{i} \rightarrow \psi_{j} \psi_{j}} = \frac{M_{h_i} \alpha^2_{jji}}{16 \pi}
\biggl( 1 - \frac{4 M^2_{\psi_{i}}}{M^2_{h_{i}}} \biggr)^{3/2}\,,
\end{eqnarray}
where the coupling constants $\alpha_{jji}$ ($j=1,2$ and $i=1,2,3$) are given in 
Eq. (\ref{psipsihi}).

\item The decay width for the $A \rightarrow \psi_{i} \psi_{i}$, $i = 1,2$, can be expressed
as
\begin{eqnarray}
\Gamma_{A \rightarrow \psi_{i}\psi_{i}} = \frac{\alpha^2_{iiA} M_{A}}{8 \pi}
\sqrt{1 - \frac{4 M^2_{\psi_{i}}}{M^2_{A}}}
\end{eqnarray}
where the vertex factors are given in Eq. (\ref{psipsihi}).
\end{itemize}

\end{itemize}

\end{document}